\documentclass[aps,10pt,prc,notitlepage,tightenlines,twocolumn,twoside,showpacs,nofootinbib,superscriptaddress,amssymb,amsmath]{revtex4-1}
\usepackage{graphicx,amsmath,amssymb}
\usepackage{hyperref}
\usepackage{breakurl}
\usepackage{color}
\usepackage{verbatim}
\usepackage{multirow}
\usepackage{tabularx}
\usepackage{slashed}
\usepackage{soul}
\usepackage{bm} 

\newcommand{\eg}{\textit{e.g.}}

\newcommand{\be}{\begin {equation}}
\newcommand{\ee}{\end{equation}}
\newcommand{\bi}{\begin{itemize}}
\newcommand{\ei}{\end{itemize}}
\newcommand{\bea}{\begin {eqnarray}}
\newcommand{\eea}{\end{eqnarray}}

\newcommand{\dif}{\mathrm{d}}

\begin{document}

\title{Heavy quarkonia production at LHC and EIC using Basis Light-Front Quantization wavefunctions}

\author{Guangyao Chen}  
\email{gchen11@gsu.edu}
\affiliation{Department of Physics and Astronomy, Iowa State University, Ames, Iowa 50011, USA}
\affiliation{Department of Physical Sciences, Perimeter College, Georgia State University, Alpharetta, Georgia 30022, USA}

\author{Yang Li}
\email{leeyoung@iastate.edu}
\affiliation{Department of Physics and Astronomy, Iowa State University, Ames, Iowa 50011, USA}


\author{Kirill Tuchin}
\email{tuchin@iastate.edu}
\affiliation{Department of Physics and Astronomy, Iowa State University, Ames, Iowa 50011, USA}

\author{James P. Vary}
\email{jvary@iastate.edu}
\affiliation{Department of Physics and Astronomy, Iowa State University, Ames, Iowa 50011, USA}

\date{\today}

\begin{abstract}
We study exclusive charmonium and bottomonium production in ultra-peripheral heavy-ion collisions and electron-ion collisions within the dipole picture. We employ heavy quarkonium light-front wavefunctions obtained within the basis light-front quantization framework, which has some features of the light-front holographic QCD. We focus on comparison with measurements of exclusive charmonium and bottomonium production in ultraperipheral $pp$, $p$Pb and PbPb collisions at LHC and find reasonable agreement with the cross sections. We also discuss the coherent production cross-section ratios of excited states to the ground state for charmonium and bottomonium, which exhibit insensitivity to the dipole model parameters. We show that electron-ion collisions provide an opportunity for a quantitative study of heavy quarkonium structure. 
\end{abstract}


\maketitle

\section{Introduction}
\label{sec_intro}

Several fundamental aspects of Quantum Chromodynamics (QCD) in the high energy limit can be understood by studying diffractive processes in high energy nuclear collisions \cite{Gribov:1984tu,Derrick:1993xh, Ahmed:1994nw}. In particular, diffractive vector meson production \cite{Ivanov:2004ax,Wolf:2009jm} provides valuable insights on gluon saturation \cite{JalilianMarian:1996xn,Gelis:2010nm,Golec-Biernat:1998js,Levin:2000mv,Gotsman:2001ne,Kowalski:2003hm,Iancu:2003ge} at low Bjorken-$x$, where gluon recombination effects become important. Moreover, generalized parton distributions (GPDs) of the target nucleus can be extracted from suitable exclusive scattering processes in terms of the squared momentum transfer $t$, which also provide the transverse spatial distribution of the nuclear partons. 

Gluon saturation is favored by data collected at various experiments at the Hadron-Electron Ring Accelerator (HERA) at DESY, the Relativistic Heavy Ion Collider (RHIC) at BNL and the Large Hadron Collider (LHC) at CERN. However, models without saturation can provide alternative descriptions of the data, see Ref.~\cite{Mantysaari:2018nng} and references therein. Until now, unambiguous evidence of saturation is lacking, due to the limited kinematic coverage of previous experiments.

Our knowledge of nucleon structure, including the saturation mechanism, has grown significantly over the past decades by colliding leptons with protons, in particularly by measuring the inclusive Deep Inelastic Scattering (DIS) at HERA \cite{Abramowicz:2015mha}. Although hadron production in semi-inclusive DIS (SIDIS) and hard exclusive processes in DIS have led to fascinating new insights into the structure of the nucleon, the current experimental information we have on nucleon structure is basically one-dimensional, that is, solely in the longitudinal direction.

Experimental evidence for gluon saturation and three-dimensional tomographic imaging of the nucleon's structure are anticipated at future experimental facilities, e.g., the Large Hadron electron Collider (LHeC) \cite{AbelleiraFernandez:2012cc} and the Electron-Ion Collider (EIC) \cite{Accardi:2012qut}. Hard exclusive processes are anticipated to play an important role in these quests. On the one hand, the luminosity at future electron ion collision facilities is expected to increase by at least two orders of magnitude over past and existing lepton-hadron collision facilities. On the other hand, it is expected that the total diffractive cross section would constitute around $30$\% of the total cross section, very close to the black disk limit of $50$\%. Consequently, the statistical uncertainty associated with the diffractive processes would be significantly reduced by the large number of events. Furthermore, several accurate measurements which are not possible at HERA would be viable at a future LHeC and at the EIC. For instance, one may anticipate measuring diffractive production of higher excited vector meson states. Such experiments could provide major new tests of meson and nucleon structures.

In the dipole picture \cite{Mueller:1989st, Kopeliovich:1993pw, Nikolaev:1990ja}, diffractive vector meson production is calculated by convoluting the photon light-front wavefunction (LFWF) and the vector meson (VM) LFWF with the dipole cross section. In the leading Fock sector, a dipole pair consists of a quark and an antiquark. This seemingly simple quantum mechanical formalism is promising for incorporating up to next-to-leading-order (NLO) corrections in QCD in the near future \cite{Lappi:2016fmu,Beuf:2016wdz}, due to the utility of the Eikonal approximation at high energy \cite{Nemchik:1994fp}. 

One of the major theoretical challenges in calculating diffractive vector meson production in the dipole picture is our poor knowledge of the LFWF of vector mesons, especially their excited states. Some of the popular vector meson LFWFs currently employed in calculating diffractive vector meson production are based on analogy with the virtual photon LFWF, such as the boosted Gaussian \cite{Brodsky:1980vj,Nemchik:1996cw} and holographic LFWFs \cite{Forshaw:2012im}. The description of higher excited states is missing within the holographic approach, while boosted Gaussian requires additional assumptions which inevitably introduce more uncertainty relative to ground states \cite{Cox:2009ag}. Since the higher excited vector meson states should have a more complicated structure than the ground state, and they are expected to be available for precise measurements at future electron-ion collision facilities, a better understanding of the vector meson sector is in order.         

The basis light-front quantization (BLFQ) approach  \cite{Honkanen:2010rc,Vary:2009gt,Zhao:2014xaa,Wiecki:2014ola,Vary:2016emi} provides a novel numerical solution for vector meson LFWFs. The heavy quarkonium system is solved using an effective Hamiltonian which has some features of the light-front holographic QCD effective Hamiltonian \cite{deTeramond:2005su,Brodsky:2014yha}. In addition to the soft-wall confinement in the transverse direction, it includes longitudinal confinement to complete the confining potential for the heavy flavors, and the one-gluon exchange interaction is implemented to generate the spin structure of heavy quarkonium \cite{Li:2015zda}. Recently, the running coupling has been implemented for the one-gluon exchange interaction and the resulting spectroscopy is improved compared with previous results \cite{Li:2017mlw}.

The LFWFs obtained within the BLFQ formalism provide a reasonable description of heavy quarkonia, including all states below the open flavor thresholds. For example, the LFWFs yield results for the mass spectroscopy, the decay constants, the r.m.s. radii \cite{Li:2015zda,Li:2017mlw} and the decay via magnetic dipole radiation \cite{Li:2018uif}. Furthermore, the BLFQ LFWFs calculated in Ref.~\cite{Li:2015zda} were employed in the dipole picture, and the results were found to be consistent with diffractive charmonium production data at HERA and the PbPb collisions at LHC \cite{Chen:2016dlk}. In particularly, we found that the diffractive production cross-section ratio $\sigma_{\Psi(2s)}/\sigma_{J/\Psi}$ exhibits significant independence of model parameters, especially for the deeply virtual processes.   

In this paper, we calculate diffractive production of vector charmonium and bottomonium states below the open flavor thresholds in the dipole picture using the LFWFs obtained in Ref.~\cite{Li:2015zda}, to be consistent with an earlier application of BLFQ LFWFs, which was primarily focused on predicting charmonium production at HERA. Diffractive production of vector charmonium and bottomonium using the LFWFs obtained in Ref.~\cite{Li:2017mlw} differs from the prediction we present here by less than $2\%$ and will be reported elsewhere.  We emphasize comparing our calculation to experimental data for ultra-peripheral $pp$, $p$Pb and PbPb collisions at LHC, which complement our previous calculations in Ref.~\cite{Chen:2016dlk}. We also make predictions for future EIC experiments for the bottomonia production.

This paper is organized as follows. In Sec.~\ref{sec:bg}, we review the vector meson production in the dipole picture and the heavy quarkonium LFWFs in the BLFQ approach. We compare the predictions of the diffractive production cross sections to the experimental data collected at LHC, including data from LHC run 2, in Sec.~\ref{sec:LHC}. We make predictions for the cross-section ratios of $\Upsilon$ production at future EIC experiments in Sec.~\ref{sec:ratio}. We summarize our results in Sec. \ref{sec:conclusion}.

\section{Background}
\label{sec:bg}

\subsection{The dipole model}
\label{ssec:dipole}

The diffractive production of vector mesons and semi-inclusive DIS can be described simultaneously in the color dipole picture, e.g., in Refs.~\cite{Golec-Biernat:1998js,Kowalski:2003hm,Rezaeian:2013tka}. In the dipole picture, due to the time dilation of the photon LFWF in the proton rest frame, both diffractive and semi-inclusive DIS can be assumed to occur in three subprocesses: first the incoming virtual photon fluctuates into a quark-antiquark pair, then the $q\bar q$ pair scatters off the proton, and finally the $q\bar q$ pair recombines to form a virtual photon or a vector meson. The effectiveness of the dipole picture relies on the separation of timescales: the lifetime of the $q\bar q$ pair at small $x$ is much longer than its typical interaction time with the target. 

In the dipole picture, the total DIS cross section can be factorized in the following form, 
\begin{eqnarray}
  && \sigma_{T,L}^{\gamma^* p}(x,Q)  = \nonumber \\ 
   \sum_{f} \int\!\dif^2\bm{r} && \int_0^1\!\frac{\dif{z}}{4\pi} 
\;(\Psi^{*}\Psi)^f_{T,L} (r,z,Q)\; \sigma_{q\bar q} (x,r) \; ,
  \label{eq:sigmaDIS}
\end{eqnarray}
with the summation over quark flavor $f$, and $(\Psi^{*}\Psi)^f_{T,L}$ denotes the overlap of the incoming and outgoing virtual photon LFWFs of the leading quark-antiquark Fock sector in the transverse (T) or longitudinal (L) polarization configurations; 
$\sigma_{q\bar q} (x,r)$ is the cross section of a $q\bar q$ pair scattering off a proton; and $Q^2=-q^2$ denotes the virtuality of the photon where $q$ represents the $4$-momentum of the photon, $\bm {r}$ is the transverse separation of the quark and antiquark, and $z$ is the LF longitudinal momentum fraction of the quark. The momentum fraction of gluons in the proton which are interacting with the dipole is specified by Bjorken-$x$. 

The diffractive vector meson production cross section can be calculated in an approach similar to the total DIS cross section. The prodcution amplitude can be calculated as \cite{Kowalski:2006hc}
\begin{eqnarray}
   && \mathcal{A}^{\gamma^* p\rightarrow Ep}_{T,L}(x,Q,\Delta)  = \nonumber \\ 
  && \mathrm{i}\,\int\!\dif^2\bm{r}\int_0^1\!\frac{\dif{z}}{4\pi}\int\!\dif^2\bm{b} 
\;(\Psi_{E}^{*}\Psi)_{T,L} (r,z,Q)  \;    \nonumber \\
&& \times    \mathrm{e}^{-\mathrm{i}[\bm{b}-(1-z)\bm{r}]\cdot\bm{\Delta}}
  \;\frac {\dif\sigma_{q\bar q}}{\dif^2 \bm b} (x,r)  \; ,
  \label{eq:newampvecm}
\end{eqnarray}
where $t= - \bm{\Delta}^2$ denotes the momentum transfer between the dipole and the nucleus. On the right-hand side, $\bm{r}$ is the transverse size of the color
dipole and $\bm{b}$ is the impact parameter of the dipole relative to the proton. $\Psi$ and $\Psi_{E}^{*}$ are LFWFs of the virtual photon and the exclusively produced vector meson respectively. The cross section then is calculated from the amplitude via 
\begin{eqnarray}
\frac{\dif \sigma^{\gamma^* p\rightarrow Ep}_{T,L}}{\dif t} = \frac{1}{16 \pi} \vert \mathcal{A}^{\gamma^* p\rightarrow
Ep}_{T,L}(x,Q,\Delta)  \vert^2 \; .
\end{eqnarray}

The photon LFWFs, which describe the amplitude for the photon to fluctuate into a quark-antiquark dipole, are usually calculated using perturbative methods  \cite{Lepage:1980fj,Forshaw:2003ki}. In the leading order (LO) of $\alpha_\text{em}$ (the fine structure constant), the normalized photon wave function for the longitudinal photon polarization, with $\lambda=0$, is given by
\begin{equation}
  \Psi_{h\bar{h},\lambda=0}(r,z,Q) =  e_f e \, \sqrt{N_c}\, 
  \delta_{h,-\bar h} \, 2Qz(1-z)\, \frac{K_0(\epsilon r)}{2\pi},   \; ,
  \label{lspinphot}
\end{equation}
and for the transverse photon polarizations, $\lambda = \pm 1$, are given by
\begin{eqnarray}
  \Psi_{h\bar{h},\lambda=\pm 1}(r,z,Q) = &&
  \pm e_f e \, \sqrt{2N_c}\,  
  \{
  \mathrm{i}e^{\pm \mathrm{i}\theta_r}[
    z\delta_{h,\pm}\delta_{\bar h,\mp} \nonumber \\  && - 
    (1-z)\delta_{h,\mp}\delta_{\bar h,\pm}] \partial_r \, + \, 
  m_f \delta_{h,\pm}\delta_{\bar h,\pm}
  \}\,
  \nonumber \\
  && \times  \frac{K_0(\epsilon r)}{2\pi},
  \label{tspinphot}
\end{eqnarray}
with $e=\sqrt{4\pi\alpha_{\mathrm{em}}}$ being the charge of electron and $e_f$ is quark charge number; the subscripts $h$ and $\bar h$ are the light-front helicities of the quark and the antiquark, respectively; and $\theta_r$ is the azimuthal angle between the vector $\bm {r}$ and the polarization $x$-axis in the transverse plane.  $K_0$ is a modified Bessel function of the second kind; and $\epsilon^2 \equiv z(1-z)Q^2+m_f^2$ and $N_c=3$ is the number of colors. We employ the LO photon LFWFs in this paper. However it is worth mentioning that promising progresses has been made in generalizing the dipole factorization for DIS at NLO \cite{Lappi:2016fmu,Beuf:2016wdz}.  

The interactions between the quark-antiquark dipole and the proton are encoded by the dipole cross section $\sigma_{q\bar q} (x,r)$. Several successful models have been proposed and the parameters were determined by fitting the theoretical total DIS cross section calculated for the proton structure function $F_2$, measured primarily at HERA. In pioneering studies of the dipole model \cite{Nikolaev:1994rd,Mueller:1994jq}, the contribution from the multi-gluon exchange diagrams were resummed up to the leading log in $1/x$, through the Balitsky-Fadin-Kuraev-Lipatov (BFKL) equation \cite{BFKL}. Later the Golec-Biernat and W\"usthoff dipole model, which is based on the scaling of saturation scale $Q_s$ as a function of Bjorken-$x$, was able to describe the total inclusive and diffractive DIS cross sections at HERA, except for the large $Q^2$ data. 

Dipole models with explicit impact parameter dependence are favored in the study of diffractive vector meson production. For instance, the diffractive vector meson production as a function of the momentum transfer between the dipole and the proton can provide valuable information on the GPDs and spatial parton distributions of the proton. The dipole cross section can be obtained by integrating over the impact parameter dependent dipole cross section as follows,
\begin{eqnarray}
\sigma_{q\bar q} (x,r)  = && \int \dif^2\bm{b} \frac {\dif\sigma_{q\bar q}(x,r,\bm{b})}{\dif^2\bm{b}}  \; .
\end{eqnarray}
In the following we discuss the bCGC and bSat dipole models since they are currently widely used in investigations of DIS and we adopt these two dipole models for our investigation.     

For small-$x$ gluons, non-linear dynamics are dominant in the saturation regime. Consequently the Balitsky-Kovchegov (BK) equation \cite{BK} is considered to be more relevant in DIS involving small-$x$ gluons. Inspired by the color glass condensate, the effective theory of QCD in the saturation regime, Iancu, Itakura, and Munier proposed the CGC dipole model, which adopts the solutions of the BFKL equation for small dipoles and the Levin–Tuchin solution \cite{Levin:1999mw} of the BK equation for larger dipoles. The CGC dipole model was generalized to bCGC dipole model by Watt and Kowalski \cite{Watt:2007nr}. The dipole proton cross section is    
\begin{eqnarray}
\label{eq:bcgc}
 && \frac {\dif\sigma_{q\bar q}(x,r,\bm{b})}{\dif^2\bm{b}} \nonumber \\
  =&& 2
  \begin{cases}
    \mathcal{N}_0 \left(\frac{rQ_s}{2}\right)^{2 \gamma_\text{eff}}  & :\quad rQ_s\le 2\\
    1-\mathrm{e}^{-\mathcal{A} \ln^2(\mathcal{B} rQ_s)} & :\quad rQ_s>2
  \end{cases},
\end{eqnarray}
where $\mathcal{N}_0$, $\mathcal{A}$ and $\mathcal{B}$ are determined such that the above equation is smooth at $rQ_s = 2$, with
\begin{equation}
\gamma_\text{eff}= \gamma_s + \frac{1}{\kappa_s \lambda_s \ln (1/x) \ln \frac{2}{rQ_s} } \; ,
\end{equation}
and the impact parameter dependence was introduced in Ref.~\cite{Watt:2007nr}, which we follow here,
\begin{equation}
Q_s(x,\bm{b}) = (\frac{x_0}{x})^{\lambda_s/2}  \exp\big(- \frac{b^2}{4 \gamma_s B_\text{CGC}}\big)~\text{GeV} \; ,
\end{equation}
where $x_0$, $\gamma_s$, $\kappa_s$, $\lambda_s$ and $B_\text{CGC}$ are parameters to be determined by inclusive DIS data \cite{Abramowicz:2015mha}. In this investigation, we use recent parametrizations by Rezaeian and Schmidt \cite{Rezaeian:2013tka} obtained by fitting to the updated combined DIS data by the ZEUS and H1 collaborations \cite{Abramowicz:2015mha}. We follow the prescription in Refs.~\cite{Watt:2007nr,Rezaeian:2013tka} for the skewedness
correction in the bCGC dipole model. $R_\text{bCGC}$ is assumed to be,
\begin{eqnarray}
  R_\text{bCGC}(\delta_\text{bCGC}) = \frac{2^{2\delta_\text{bCGC}+3}}{\sqrt{\pi}}
\frac{\Gamma(\delta_\text{bCGC}+5/2)}{\Gamma(\delta_\text{bCGC}+4)}  \; , 
 \label{eq:Rg_bCGC}
\end{eqnarray}
with
\begin{equation}
\delta_\text{bCGC}  \equiv
\frac{\partial\ln\left(\mathcal{A}_{T,L}^{\gamma^* p\rightarrow Ep}\right)}{\partial\ln(1/x)} \; ,
\end{equation}
where $\mathcal{A}_{T,L}$ is the production amplitude in Eq.~(\ref{eq:newampvecm}). The obtained $R_\text{bCGC}$ is then applied multiplicatively to the production amplitude.

The bSat dipole model was first proposed by Kowalski and Teaney \cite{Kowalski:2003hm} based on the Glauber-Mueller formula \cite{Mueller:1989st} and assumes the dipole cross section as follows:
\begin{align}
\label{eq:bsat}
  \frac{\dif\sigma_{q\bar{q}}}{\dif^2\bm{b}} = 2\left[1-\exp\left(-\frac{\pi^2}{2N_c}r^2\alpha_s(\mu^2)xg(x,\mu^2)T(b)\right)\right] \; ,
\end{align}
where $\alpha_s$ is determined using LO evolution of the running coupling, with fixed number of flavors $N_f$. $\mu^2$ is related to the
dipole size $r$ through $\mu^2=4/r^2+\mu_0^2$. The gluon density is determined using LO Dokshitzer-Gribov-Lipatov-Altarelli-Parisi evolution \cite{Bartels:2002cj} from an initial scale $\mu_0^2$, where the initial gluon density is,
\begin{eqnarray}
 \label{eq:inputgluon}
  xg(x,\mu_0^2) = A_g\,x^{-\lambda_g}\,(1-x)^{5.6}. 
 \end{eqnarray} 
The impact parameter dependence was introduced through the proton shape function,
\begin{equation}
T_G(b) =\exp(-b^2/2B_G)/(2\pi B_G)   \;  ,
\end{equation} 
with $B_G = 4$~GeV$^{-2}$. In the bSat dipole model, $\mu_0$, $A_g$ and $\lambda_g$ are parameters to be determined by the inclusive DIS data \cite{Abramowicz:2015mha}. We use values for these parameters
given in Ref.~\cite{Rezaeian:2012ji} for this investigation. We follow the
prescription in Ref.~\cite{Kowalski:2006hc,Rezaeian:2012ji} for the skewedness correction in the bSat dipole model. $R_\text{bSat}$ is assumed to be
\begin{eqnarray}
  R_\text{bSat}(\delta_{bSat}) = \frac{2^{2\delta_\text{bSat}+3}}{\sqrt{\pi}}
\frac{\Gamma(\delta_\text{bSat}+5/2)}{\Gamma(\delta_\text{bSat}+4)} \; ,
 \label{eq:Rg_bsat}
\end{eqnarray}
with
\begin{equation}
\delta_\text{bSat}  \equiv
\frac{\partial\ln\left[xg(x,\mu^2)\right]}{\partial\ln(1/x)}.
\end{equation}
The obtained $R_\text{bSat}$ is then applied multiplicatively to the gluon density function in Eq.~(\ref{eq:bsat}). 

In order to make comparison with experimental data, it is a widely adopted practice to include another phenomenological correction in the calculation, e.g., the contribution from the real part of the scattering amplitude is conventionally incorporated by multiplying the cross section by a factor
$(1+\beta^2)$~\cite{Kowalski:2006hc}, where $\beta$ is the ratio of the real and imaginary parts of the scattering amplitude, and is calculated as \cite{Ryskin:1995hz}
\begin{eqnarray}
  \beta = \tan(\pi\lambda/2), \quad\text{with}\quad \lambda \equiv \frac{\partial\ln\left(\mathcal{A}_{T,L}^{\gamma^* p\rightarrow Ep}
\right)}{\partial\ln(1/x)}  \; ,
  \label{eq:beta}
\end{eqnarray}
where $\mathcal{A}_{T,L}$ is the production amplitude in Eq.~(\ref{eq:newampvecm}).

\subsection{Heavy quarkonium in the basis approach}
\label{ssec:BLFQ}

The understanding of the meson structure from first principles is hindered by unsolved problems in the non-perturbative regime of QCD such as the origin of confinement and the dynamics of chiral symmetry breaking. However, a quantitative vector meson LFWF is critical for the study of diffractive vector meson production in the dipole picture. Until now, the boosted Gaussian (bG) LFWF \cite{Brodsky:1980vj, Forshaw:2003ki} and the holographic QCD LFWF \cite{deTeramond:2005su,Brodsky:2014yha} are two popular choices.   

Both the bG and holographic LFWFs rely on the assumption that the vector meson LFWFs have the same spin structure as a virtual photon. They differ only by the specifications of the scalar components of the LFWFs. For example, the bG LFWFs are obtained by boosting a Gaussian type wavefunction in the meson rest frame to the infinite momentum frame, and the holographic LFWFs are obtained by solving a Schr\"odinger-like eigen equation with an effective light-front holographic Hamiltonian \cite{deTeramond:2005su}. Both of them have enjoyed successes in phenomenological applications. Nevertheless, there are several limitations to be overcome. The major drawback of the bG LFWFs is that the parameters cannot be uniquely determined which introduces uncertainties to the calculation. On the other hand, the holographic LFWFs have better control over the model parameters, but its application is limited to the massless or small quark mass regime. Moreover, there are some ambiguities in the description of the higher excited states within the light-front holographic QCD approach. In the following we discuss the holographic LFWF and sketch how the BLFQ formalism extends the Hamiltonian to include additional features of QCD.      

In principle, the hadron mass spectrum and the associated light-front amplitudes can be obtained by solving the eigen equation of the light-front QCD (LFQCD) Hamiltonian operator $H_\textsc{lfqcd}$,
\begin{equation}
 H_\textsc{lfqcd}|\psi_h\rangle = M^2_h|\psi_h\rangle \; .
\end{equation}    
However, it is a formidable task to work with the QCD Hamiltonian directly. As a compromise for the quarkonium system, Brodsky and de Teramond work with the semi-classical approximation of light-front QCD, the light-front holographic QCD \cite{deTeramond:2005su}, which is based on the correspondence between anti-de Sitter (AdS) space and QCD,
\begin{equation}
 H_\text{holographic} \equiv  \frac{\bm k^2 + m_q^2}{z(1-z)}  + \kappa_\text{con}^4 \bm \zeta_\perp^2  \; , 
 \label{eq:holographic}
\end{equation}  
where $\bm k$ and $z$ are the transverse momentum and longitudinal momentum fraction of the quark, respectively; $\bm \zeta_\perp \equiv \sqrt{z(1-z)} \bm r$ is the transverse separation of quark and antiquark on the light front; and $\kappa_\text{con}$ is the strength of the confinement. The meson mass spectrum can be reproduced well by solving Eq.~(\ref{eq:holographic}) \cite{deTeramond:2005su} and the associated meson LFWFs yield diffractive $\rho$ and $\phi$ production which are consistent with measurements at HERA and LHC \cite{Forshaw:2012im}. The light-front holographic QCD was derived in the massless limit. It can be extended to small quark masses using the invariant mass ansatz \cite{Brodsky:2008pg}. It is a challenge to describe the heavy quarkonium system using light-front holographic Hamiltonian in Eq.~(\ref{eq:holographic}). Even within the light quark sector, it is a challenge to include higher excited states within the light-front holographic formalism only. 

The BLFQ approach to heavy quarkonia \cite{Li:2015zda} transcends the above limitations and generalizes the holographic QCD to heavy flavor sector by introducing a longitudinal potential and including the the one-gluon exchange dynamics \cite{Li:2015zda},   
\begin{eqnarray}
\label{eq:Ham}
 H_\text{BLFQ} = H_\text{holographic} 
 - \frac{\kappa_\text{con}^4}{4 m_q^2}\partial_z \big(z(1-z) \partial_z \big)   \nonumber  \\
 -\frac{4\pi C_F \alpha_s}{Q^2} \bar u_{s}(k)\gamma_\mu u_{s'}(k') \bar v_{\bar s'}(\bar k') \gamma^\mu v_{\bar s}(\bar k)\;,
 \end{eqnarray} 
where $C_F=\frac{4}{3}$, $Q^2=-\frac{1}{2}(k'-k)^2-\frac{1}{2}(\bar k-\bar k')^2$. This BLFQ approach is suitable for generating all the states of the heavy quarkonia systems governed by the same Hamilonian in the chosen Fock space representation. The second term, which is the longitudinal confining potential, can be solved analytically and the resulting wavefunction resembles the known asymptotic parton distribution $\phi^\textsc{da}(x)\sim x^\alpha(1-x)^\beta$. The last term, the one-gluon exchange interaction which is derived from light-front QCD, provides the short-distance physics and spin structures needed for the angular excitations and the hyperfine structure. With this effective Hamiltonian, there is no need for additional assumptions about the spin structure of the bound states, including all the excited states. In particular, the one-gluon exchange interaction gives rise to D-wave components in our vector meson LFWFs.

The model for the BLFQ effective Hamiltonian has several parameters. The strong coupling constant $\alpha_s$, the effective quark mass $m_q$ and the confining strength $\kappa_\text{con}$. These parameters are determined by fitting the mass spectrum of the Hamiltonian to the experimental spectrum for heavy quarkonium states below the open-flavor thresholds. There are $8$ charmonium states ($2$ of which are vector mesons) and $14$ bottomonium states ($4$ of which are vector mesons), that fall into this category. In an initial study \cite{Li:2015zda}, $\alpha_s$ is fixed, with $\alpha_s(M_{c\bar c}) \simeq 0.36$ and  $\alpha_s(M_{b\bar b}) \simeq 0.25$. They are related through the pQCD evolution of the coupling constant. With a root-mean-square (r.m.s.) deviation in their masses from experiment of about $50$~MeV, the other two parameters are fitted to be $m_c = 1.522$~GeV, $\kappa_{con} = 0.938$~GeV for charmonium and  $m_b = 4.763$~GeV and $\kappa_{con} = 1.490$~GeV for bottomonium. In a subsequent investigation \cite{Li:2017mlw}, the evolution of the strong coupling $\alpha_s$ as a function of invariant 4-momentum transfer is included. With a resulting r.m.s. deviation in their masses from experiment of about $31$~MeV, the other two parameters are fitted to be $m_c = 1.603$~GeV, $\kappa_{con} = 0.966$~GeV for charmonium. For the bottomonium states, the fitting gives $m_b = 4.902$~GeV and $\kappa_{con} = 1.389$~GeV, with an r.m.s. deviation in their masses from experiment of about $38$~MeV. The resulting LFWFs from both investigations \cite{Li:2015zda,Li:2017mlw} predict the decay constants, the form factors and the charge radii which compare reasonably well to the experiments and other established methods, such as Lattice QCD and the Dyson-Schwinger Equation. The resulting LFWFs from  Ref.~\cite{Li:2015zda} were also employed for the calculation of diffractive charmonium production in the dipole picture, without adjusting the parameters \cite{Chen:2016dlk}. The charmonium cross section predicted by the LFWFs from Ref.~\cite{Li:2015zda} is in reasonable agreement with experimental data at HERA, RHIC and LHC.

The appealing features of heavy quarkonium LFWFs obtained from the BLFQ approach can be summarized as follows. First, the BLFQ formalism provides a unified description for a variety of observables, such as mass spectroscopy and decay constants. The BLFQ LFWFs also provide valuable insights for additional quantities such as the form factors and the charge radii. Second, the excited states are described without introducing additional assumptions, e.g., the LFWFs for all charmonium and bottomonium states below the open-flavor thresholds are dictated by the effective Hamiltonian in Eq.~(\ref{eq:Ham}). Predictions for excited states, from BLFQ and from other approaches, will be valuable for the future EIC where these states can be accurately measured. It is also anticipated that we would be able to study the dipole-nucleus interaction in the time-dependent BLFQ framework \cite{Zhao:2013cma, Chen:2017uuq}, and provide a more consistent description of the diffractive vector meson procution process.

In our previous calculation \cite{Chen:2016dlk} and also in this investigation, we employ LFWFs calculated in Ref.~\cite{Li:2015zda} to study diffractive heavy quarkonium production in various experiments. The quark masses obtained by the fitting were regarded as the effective quark masses in the bound states, which are not necessarily the same as the quark masses in the virtual photon LFWF or the dipole cross section. We hypothesize that the energy scales of QCD interaction are different in the above processes thus effective quark masses could be different. We will specify our choice of quark mass for each of our applications. 

\subsection{Diffraction off a nucleus}
\label{ssec:BLFQ}

Accurate measurements of the diffractive events at future EIC facilities will provide a three dimensional tomographic scan of the parton distribution inside a nucleus, and will also provide invaluable insights into the gluon saturation mechanism in the small-$x$ regime \cite{AbelleiraFernandez:2012cc, Accardi:2012qut}. The EIC will provide a wide variety of heavy-ion beams, two to three orders of magnitude increase in luminosity (comparing to existing experiments) and a versatile range of kinematics for the study of diffractive processes \cite{Accardi:2012qut}. Compared to electron-proton collisions, electron-nucleus collisions reach the saturation regime at much lower energy, because the saturation effect is amplified by the number of nucleons along the path of the projectile ($\sim A^{1/3}$, with $A$ the atomic number of the nucleus). For this reason, studies of vector meson production in ultra-peripheral heavy-ion collisions (UPC), where two heavy ions scatter without overlap at large impact parameter, have provided a complementary look into the higher energy scale \cite{Baltz:2007kq}. 

It is straightforward and intuitive to extend the dipole formalism in diffractive DIS from $ep$ collision to $e$A and AA collisions. With an impact parameter dependent dipole model, the nucleus can be regarded as a collection of nucleons according to a given nuclear transverse density distribution, e.g., the Woods-Saxon distribution. If the Bjorken $x$ of a parton in the hadron is small, e.g., $x\ll A^{-1/3}/(M_NR_p)\sim 10^{-2}$ ( $M_N$ is the mass of the nucleus and $R_p$ is the proton radius), its wavelength in the $x^-$ direction is larger than the radius of the nucleus, such that the exact position of each nucleon within the nucleus is not significant. Consequently the cross section should be calculated by averaging over all possible nuclear configurations at the cross section level,
\begin{eqnarray}
    \frac{{\rm d}\sigma_{\rm total}}{{\rm d} t}=
    \frac{1}{16\pi}\left<\left|\mathcal{A}(x, Q^2, t, \Omega)\right|^2\right>_\Omega   \; ,
\end{eqnarray}
where $\Omega$ denotes nucleon configurations. 

The diffractive vector meson production can be further classified into two cases: the coherent and incoherent productions. In a coherent event, the incoming photon interacts coherently with the whole nucleus. In the case of an incoherent event, the incoming photon interacts not with the whole nucleus, but rather with a single nucleon. The coherent cross section should be calculated by averaging over all possible nuclear configurations at the amplitude level, 
\begin{eqnarray}
    \frac{{\rm d}\sigma_{\rm coherent}}{{\rm d} t}=
    \frac{1}{16\pi}\left|\left< \mathcal{A}(x, Q^2, t, \Omega)\right>_\Omega \right|^2   \; .
\end{eqnarray}
In this paper we will focus on coherent heavy quarkonium production.

One additional approximation is needed for calculating the diffractive VM production in UPC using the dipole model. In the rest frame of one of the ions, the target ion, the exclusive vector meson production can be regarded as a result of the scattering of equivalent photons radiated by the incident ion \cite{Baltz:2007kq}, and thus the VM production cross section can be calculated as
\begin{equation}
\label{eq:upc-xs}
	\sigma = \int \mathrm{d} \omega \frac{n(\omega)}{\omega} \sigma^{\gamma A}  \; ,
\end{equation}
where $\sigma^{\gamma A}$ is the photon-nucleus cross section and $n(\omega)$ is energy spectrum of the equivalent flux of photons generated by the projectile. In a modified version of the Weizs\"acker-Williams approximation, the equivalent photon flux associated with a proton is  \cite{Bertulani:2005ru}, 
\begin{eqnarray}
\frac{dN_{\gamma}(\omega)}{d\omega}  & =  & \frac{\alpha_\text{em}}{2\pi\,\omega}\, \left[1+\left(1-\frac{2\omega}{\sqrt{s}} \right)^2\right] \nonumber \\
& \times & \left(\ln \xi - \frac{11}{6} + \frac{3}{\xi}-\frac{3}{2\xi^2}+\frac{1}{3\xi^3}\right),
\label{fluxint}
\end{eqnarray}
where $\omega$ is the photon energy and $\sqrt{s}$ is the center-of-mass energy between two colliding nuclei. The dimensionless quantity $\xi$ equals $1+ (Q_0^2/Q_{\mathrm{min}}^2)$ with $Q_0^2=0.71$ GeV$^2$ and $Q_{\mathrm{min}}^2=\omega^2/\gamma_L^2$, where $\gamma_L=\sqrt{s}/(2m_p)$ is  the Lorentz factor of the projectile proton beam. The photon flux associated with a nucleus, with number of protons $Z$ and radius $R_A$, is
\begin{equation}
	n(\omega)=\frac{2Z^2 \alpha_{em}}{\pi \beta} \left[ \xi K_0(\xi) K_1(\xi) - \frac{\xi^2}{2}(K_1^2(\xi) - K_0^2(\xi) ) \right] \; ,
\end{equation}
integrated over all possible impact parameters $b>b_\text{min}=2R_A$. In the above expression, $\xi = 2\omega R_A/(\gamma \beta)$, and $\gamma$ is the Lorentz boost factor of the beam in the center of mass frame, $\beta\approx 1$ is the velocity of the projectile nucleus and $K_0$ and $K_1$ are modified Bessel functions of the second kind.  

\begin{figure}[!t]
\centering
\includegraphics[width=0.47\textwidth]{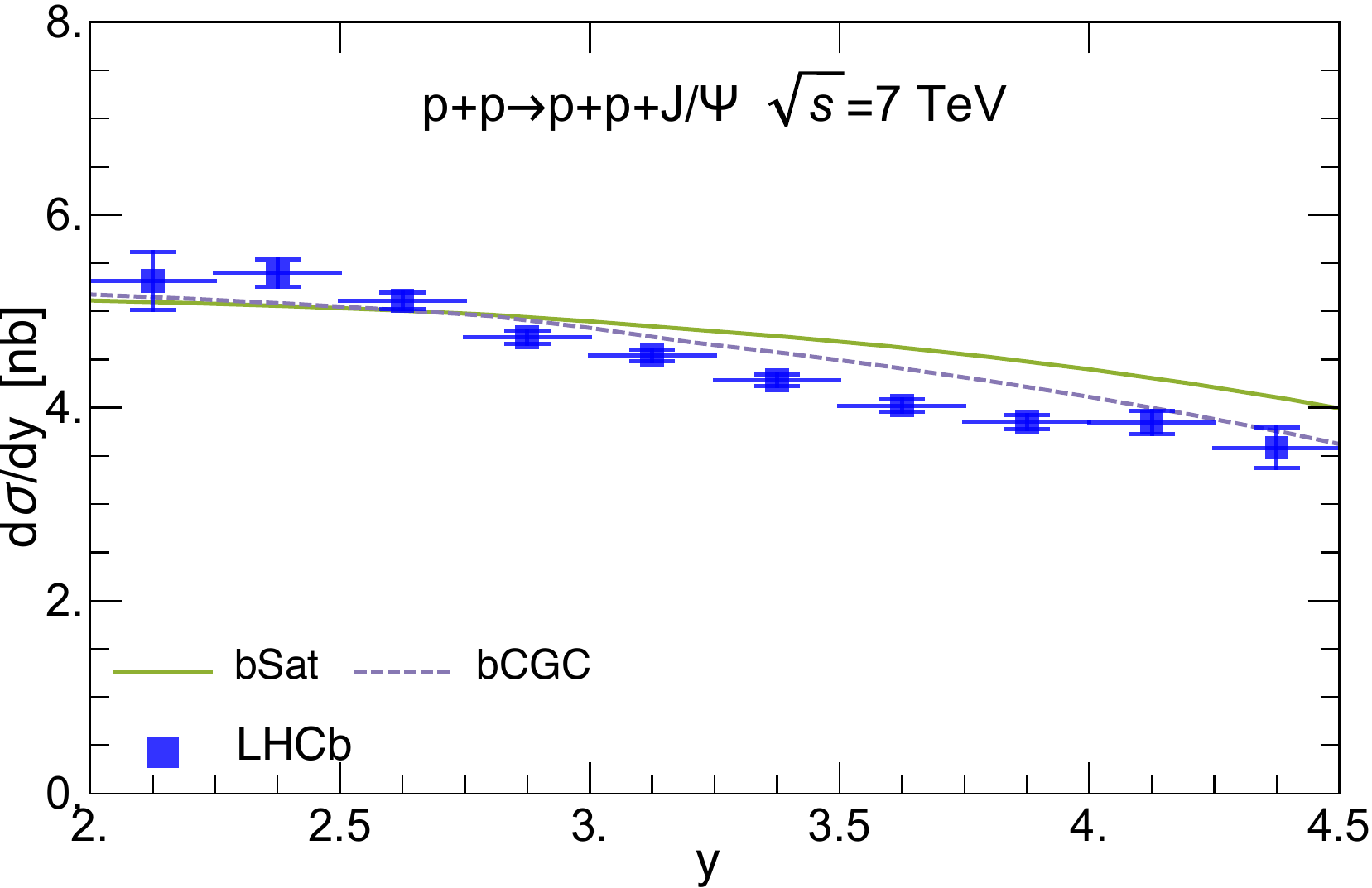}
\includegraphics[width=0.47\textwidth]{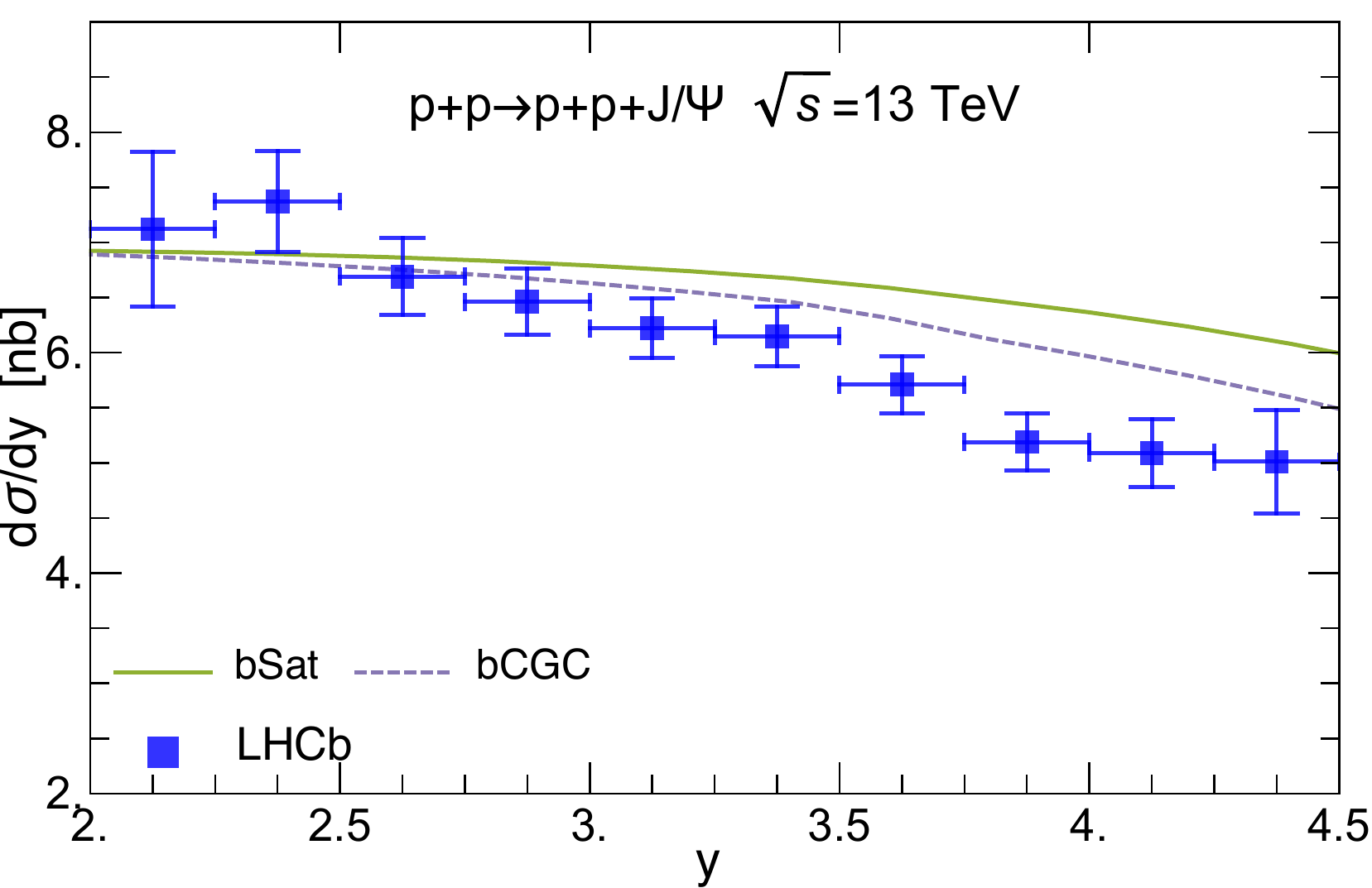}
\caption{The predictions using the BLFQ LFWF for the central exclusive $J/\Psi$ production in $pp$ collisions at center-of-mass energy $\sqrt{s}=7$~TeV (up panel), and at center-of-mass energy $\sqrt{s}=13$~TeV (bottom panel), compared with the measurements by the LHCb collaboration~\cite{Aaij:2014iea,Aaij:2018arx}.
The solid curve is produced using the bSat dipole model parametrization with $m_c =1.4$~GeV~\cite{Rezaeian:2012ji}. The dashed curve is produced using the bCGC dipole model parametrization  with $m_c =1.4$~GeV~\cite{Rezaeian:2013tka}. Error bars show the statistical and systematic uncertainties added in quadrature. 
} 
\label{pic:ppjpsi}
\end{figure}

\section{Exclusive heavy quarkonium production at the LHC}
\label{sec:LHC}

In this section, we compare the predictions using BLFQ LFWFs to measurements of exclusive charmonium and bottomonium production in ultraperipheral $pp$, $p$Pb and PbPb collisions at various energies, including recent data collected at run $2$ LHC energies. For such a purpose, we select one representative bSat dipole model parametrization from Ref.~\cite{Rezaeian:2012ji} with $m_c =1.4$~GeV, which corresponds to bSat V from Table 1 in Ref.~\cite{Chen:2016dlk}. We select one representative impact parameter dependent Color Glass Condensate dipole model (bCGC) parametrization from Ref.~\cite{Rezaeian:2013tka} with $m_c =1.4$~GeV, which corresponds to bCGC III from Table 2 in Ref.~\cite{Chen:2016dlk}. Both are fitted to the combined DIS data released in 2015 \cite{Abramowicz:2015mha}. Throughout this section, bSat and bCGC dipole model parametrizations are adopted from Ref.~\cite{Rezaeian:2012ji, Rezaeian:2013tka}. Our study complements investigations on heavy quarkonium production at LHC using other phenomenological LFWFs \cite{Goncalves:2017wgg, Kopeliovich:2001ee,Goncalves:2014wna,Xie:2016ino,Mantysaari:2017dwh,Dutta:2017kju,Carvalho:2017vtw}.

\subsection{Ultra-peripheral $pp$ collisions}
The ultra-peripheral $pp$ collision at high center-of-mass energy up to multi-TeV per nucleon can provide insight on gluons with Bjorken-x as small as $10^{-5} \sim 10^{-6}$. The LHCb collaboration has reported
$J/\Psi$ and $\Psi(2s)$ production in $pp$ collisions at a center-of-mass energy $\sqrt{s}=7$~TeV~\cite{Aaij:2014iea},
and at a center-of-mass energy $\sqrt{s}=13$~TeV~\cite{Aaij:2018arx}. The LHCb collaboration has also reported
$\Upsilon$ production in $pp$ collisions at center-of-mass energies $\sqrt{s}=7$~TeV and $\sqrt{s}=8$~TeV \cite{Aaij:2015kea}.

Heavy quarkonia are produced exclusively through the interaction between the photon emitted by one of the protons and the other proton by exchange of a colorless strongly coupled object. The large mass of the produced heavy quarkonium provides a hard scale which supports the application of the dipole model in such diffractive process. The Bjorken-$x$ of the gluons being probed through the exclusive heavy quarkonium production in the $pp$ collision is $x \approx m_V e^{\pm y} /\sqrt{s} $ where $m_V$ and $y$ are the mass and the rapidity of the produced heavy quarkonium, and $s$ the center-of-mass energy squared. The gluon distribution in the regime $10^{-2} < x < 10^{-6}$ is  constrained by experimental measurements reported in Ref.~\cite{Aaij:2014iea,Aaij:2018arx,Aaij:2015kea}. It is thus possible for us to study the heavy quarkonium production in $pp$ collisions using the dipole model and the heavy quarkonium LFWFs obtained within the BLFQ framework to provide insights on the gluon distribution.     
  
In Fig.~\ref{pic:ppjpsi}, we show the predictions using the BLFQ LFWF for the central exclusive $J/\Psi$ production in $pp$ collisions at center-of-mass energy $\sqrt{s}=7$~TeV (up panel), and at center-of-mass energy $\sqrt{s}=13$~TeV (bottom panel), compared with the measurements by the LHCb collaboration~\cite{Aaij:2014iea,Aaij:2018arx}. The solid and dashed curves are produced using the bSat dipole model parametrization and the bCGC dipole model parametrization, respectively. Using both bSat and bCGC dipole models, the predictions using the BLFQ LFWF are close to experiment for the rapidity regime $2 < y < 3$ and overestimate the $J/\Psi$ production at larger rapidities. Note that at very high rapidity, both the small and large Bjorken-x gluons in the proton contribute to the interaction: a small-$x$ photon can scatter off a large-$x$ gluon or vice versa, which add to the theoretical uncertainties for $J/\Psi$ production at large rapidity.

\begin{figure}[!t]
\centering
\includegraphics[width=0.47\textwidth]{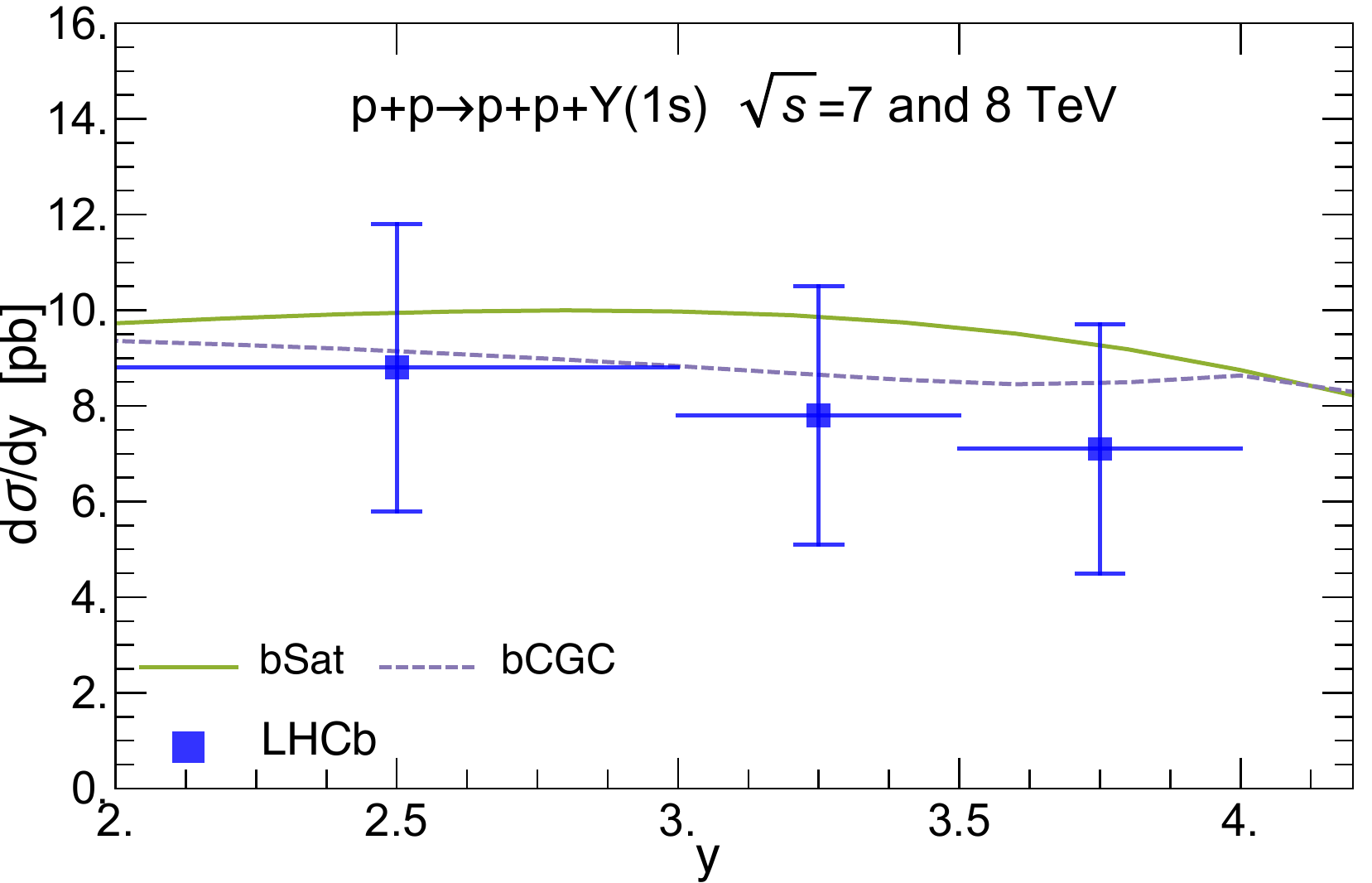}
\caption{(Colors online) the predictions using the BLFQ LFWF for the central exclusive $\Upsilon(1s)$ production in $pp$ collisions at center-of-mass energy $\sqrt{s}=7$~TeV, and $\sqrt{s}=8$~TeV, compared with the measurements by the LHCb collaboration~\cite{Aaij:2015kea}.
The solid curve is produced using the bSat dipole model parametrization with $m_c =1.4$~GeV~\cite{Rezaeian:2012ji}. The dashed curve is produced using the bCGC dipole model parametrization with $m_c =1.4$~GeV~\cite{Rezaeian:2013tka}. Error bars show the statistical and systematic uncertainties added in quadrature. 
}
\label{pic:ppupsilon}
\end{figure}

In Fig.~\ref{pic:ppupsilon}, we show the predictions using the BLFQ LFWF for the central exclusive $\Upsilon(1s)$ production in $pp$ collisions at center-of-mass energy $\sqrt{s}=7$~TeV, and $\sqrt{s}=8$~TeV, compared with the measurements by the LHCb collaboration~\cite{Aaij:2015kea}.
The solid and dashed curves are produced using the bSat dipole model parametrization and the bCGC dipole model parametrization, respectively. The yield of $\Upsilon(1s)$ is much lower compared to the charmonium production, which resulted in a larger uncertainty for the experiment data. For both bSat and bCGC dipole models, the predictions using the BLFQ LFWF agree with data within experimental uncertainty. Note that for $\Upsilon(1s)$ production, the Bjorken-x of the gluons being probed are roughly three times larger compared with the production of $J/\Psi$ at the same rapidity in $pp$ collisions at the same center-of-mass energy.

Overall, within the dipole model, the predictions of BLFQ LFWFs for the $J/\Psi$ and $\Upsilon(1s)$ agree with the latest experimental data from $pp$ collision at various energies at the LHC \cite{Aaij:2014iea,Aaij:2018arx,Aaij:2015kea}.

\subsection{Ultra-peripheral $p$Pb collisions}

\begin{figure}[!t]
\centering
\includegraphics[width=0.47\textwidth]{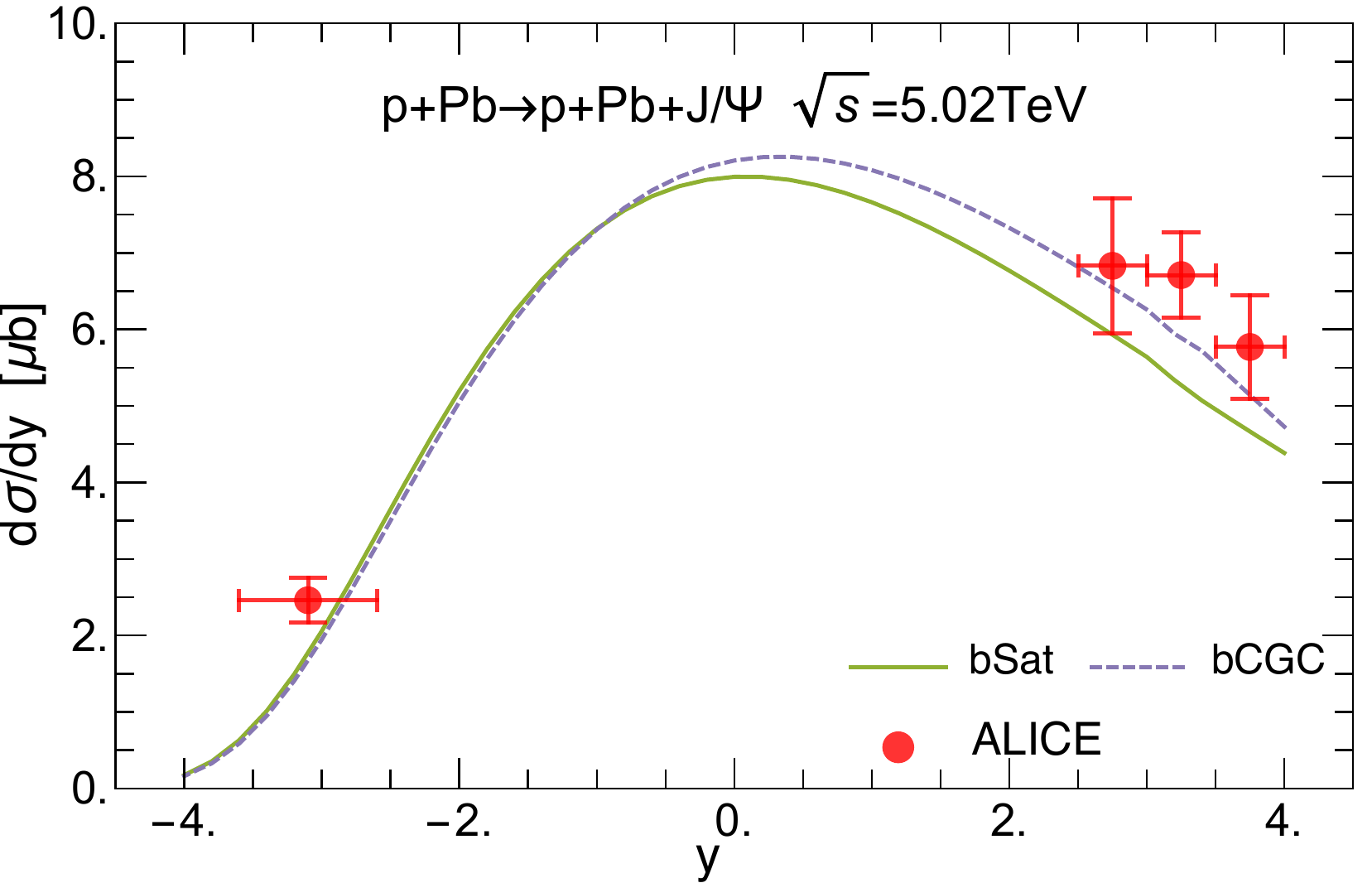}
\caption{(Colors online) the predictions using the BLFQ LFWF for the central exclusive $J/\Psi$ production in proton-lead collisions at center-of-mass energy $\sqrt{s}=5.02$~TeV, compared with the measurements by the ALICE collaboration~\cite{TheALICE:2014dwa}.
The solid curve is produced using the bSat dipole model parametrization with $m_c =1.4$~GeV~\cite{Rezaeian:2012ji}. The dashed curve is produced using the bCGC dipole model parametrization with $m_c =1.4$~GeV~\cite{Rezaeian:2013tka}. Error bars show the statistical and systematic uncertainties added in quadrature. 
}
\label{pic:pPb}
\end{figure}

In a $p$Pb collision, heavy quarkonium can be produced by the interaction of a photon with either a proton or a lead nucleus, where the photon is emitted from one of the two colliding particles. However, the density of photons emitted by the nuclear projectile is enhanced by the atomic number $Z$, so photon emission from the ion is strongly enhanced with respect to that from the proton. Consequently, the $\gamma + p \rightarrow p + V$ process strongly dominates over the process $\gamma + Pb \rightarrow Pb + V$.  

In $p$Pb collision, for example, if we set the proton motion in the $\eta<0$ direction, the $\gamma {p}$ center-of-mass energy $W_{\gamma {p}}$ is determined by the produced vector meson rapidity: $W_{\gamma {p}}^2 = 2 E_p M_V \exp (-y)$, where $M_V$ is the vector meson mass, $y$ is the vector meson rapidity and $E_p$ is the proton energy ($E_p =  4$ TeV in the lab frame). The Bjorken-$x$ of the gluons being probed is given by  $x = (M_V /W_{\gamma {p}})^2$. A unique feature of the $p$Pb asymmetric collision is that the diffractive vector meson production at different center-of-mass energies $W_{\gamma {p}}$ can be investigated simultaneously. For instance, for the exclusive $J/\Psi$ photoproduction off protons in ultra-peripheral $p$Pb collisions at $\sqrt{s}=5.02$~TeV, the $J/\Psi$ produced within the rapidity regime $2.5 < y < 4.0$ corresponds to $21<W_{\gamma {p}}<45$~GeV and the $J/\Psi$ produced within the rapidity regime $-3.6 < y < -2.6$ corresponding to $577< W_{\gamma {p}}<952$~GeV. The $J/\Psi$ produced within the rapidity regime $-3.6 < y < -2.6$ thus can provide valuable constraints to the gluon distribution at small-$x$.  

In Fig.~\ref{pic:pPb}, we show the predictions using the BLFQ LFWF for the central exclusive $J/\Psi$ production in $p$Pb collision at center-of-mass energy $\sqrt{s}=5.02$~TeV, compared with the measurements by the ALICE collaboration~\cite{TheALICE:2014dwa}.
The solid and dashed curves are produced using the bSat dipole model parametrization and the bCGC dipole model parametrization, respectively. Using the BLFQ $J/\Psi$ LFWFs, both bSat and bCGC dipole models predict a yield slightly smaller than the data. The predictions of bSat and bCGC agree with each other for $y<0$ which corresponds to high center of mass energy $W_{\gamma {p}}$, and the prediction of bSat deviates from the bCGC prediction for $y>0$ which corresponds to low center of mass energy $W_{\gamma {p}}$. Note that the data points with $y>0$ correspond to Bjorken-$x$ larger than $0.01$ in the dipole model, while the dipole model is more reliable for Bjorken-$x$ much smaller than $0.01$. Indeed we observe that both the predictions of bSat and bCGC for $y<0$ are within experimental uncertainty.

\begin{figure}[!t]
\centering
\includegraphics[width=0.47\textwidth]{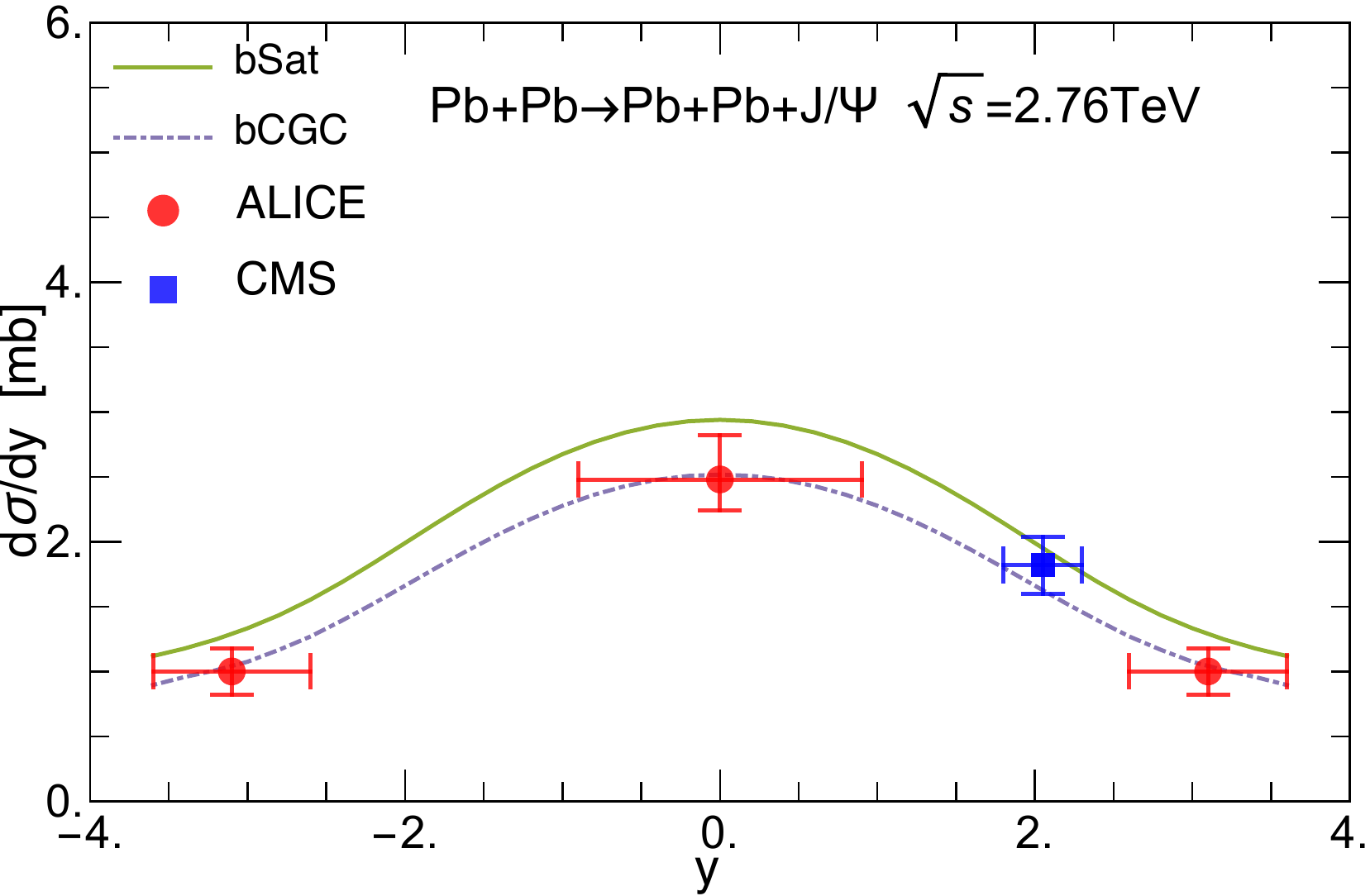}
\includegraphics[width=0.47\textwidth]{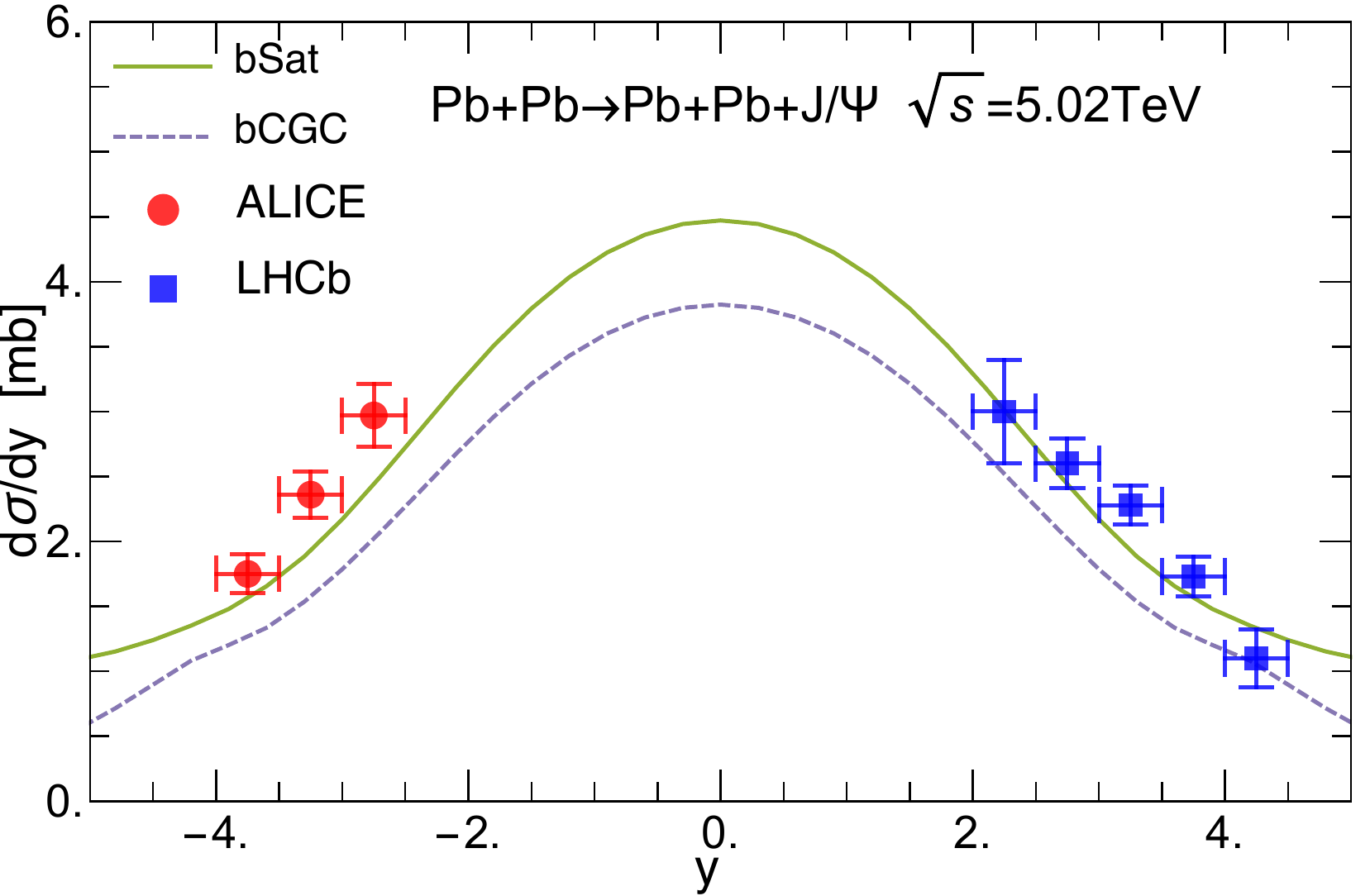}
\caption{(Colors online) the predictions using the BLFQ LFWF for the coherent $J/\Psi$ production in PbPb collisions at center-of-mass energy $\sqrt{s}=2.76$~TeV (top panel), and at center-of-mass energy $\sqrt{s}=5.02$~TeV (bottom panel), compared with the measurements by the ALICE collaboration~\cite{Abbas:2013oua,Kryshen:2017jfz}, the CMS collaboration~\cite{Khachatryan:2016qhq} and the LHCb collaboration~\cite{LHCb:2018ofh}.
The solid curve is produced using the bSat dipole model parametrization with $m_c =1.4$~GeV~\cite{Rezaeian:2012ji}. The dashed curve is produced using the bCGC dipole model parametrization with $m_c =1.4$~GeV~\cite{Rezaeian:2013tka}. Error bars show statistical uncertainties only.
}
\label{pic:PbPb}
\end{figure}

\subsection{Ultra-peripheral PbPb collisions}
Ultra-peripheral heavy ion collisions can provide photon-nucleus interactions in the kinematic regime of $Q^2 \approx 0$~GeV$^2$. Such studies thus can shed light not only on small-$x$ gluon distribution for nucleons but also for nuclei, leading to a better understanding of cold nuclear effects in high energy nuclear collisions. In ultra-peripheral PbPb collisions, the yield of heavy quarkonium is enhanced due to the large photon density ($\sim Z^2$) and the large number of target nucleons. One of the appealing facts about heavy quarkonium production in ultra-peripheral PbPb collisions is that three different collaborations have measured the $J/\Psi$ production at two different center-of-mass energies \cite{Abbas:2013oua,Kryshen:2017jfz, Khachatryan:2016qhq, LHCb:2018ofh} and the data sets are in reasonable agreement.  

The exclusive photoproduction of heavy quarkonium can be either coherent, characterized by low transverse momentum of the produced heavy quarkonium, where the photon couples coherently to almost all the nucleons; or incoherent, characterized by large transverse momentum of the produced heavy quarkonium, where the photon couples to a single nucleon. We focus on coherent heavy quarkonium production here. In Fig.~\ref{pic:PbPb}, we show the predictions using the BLFQ LFWF for the coherent $J/\Psi$ production in PbPb collisions at center-of-mass energy $\sqrt{s}=2.76$~TeV (top panel), and at center-of-mass energy $\sqrt{s}=5.02$~TeV (bottom panel), compared with the measurements by the ALICE collaboration~\cite{Abbas:2013oua,Kryshen:2017jfz}, the CMS collaboration~\cite{Khachatryan:2016qhq} and the LHCb collaboration~\cite{LHCb:2018ofh}. The solid and dashed curves are produced using the bSat dipole model parametrization and the bCGC dipole model parametrization, respectively. For the coherent $J/\Psi$ production in PbPb collisions at center-of-mass energy $\sqrt{s}=2.76$~TeV, the bSat dipole model prediction slightly overestimates the yield while the bCGC model agrees with the data. On the other hand, for the coherent $J/\Psi$ production in PbPb collisions at center-of-mass energy $\sqrt{s}=5.02$~TeV, the bSat dipole model prediction agrees with the data within uncertainty while the bCGC dipole model prediction slightly underestimates the yield. Note that the theoretical uncertainties for $J/\Psi$ production at large rapidity are larger compared to central rapidity as well.

\section{The cross-section ratio}
\label{sec:ratio}

In a previous investigation \cite{Chen:2016dlk}, we studied the ratio of the $\Psi(2s)$ cross section to the $J/\Psi$ cross section as a function of $Q^2$. If we assume that the quark-antiquark pair originating from quantum fluctuation of the virtual photon scatters universally on the nuclear target for the production of different states of the same quarkonium system, the uncertainties coming from the dipole model parametrizations may partially cancel in the ratio of different states, e.g., $J/\Psi$ and $\Psi(2s)$. Under such an assumption, the cross-section ratio of higher excited states over the ground state should also exhibit weaker dependence on the dipole model than the cross section itself. Our calculation agreed with the experiment data collected at HERA \cite{Abramowicz:2016ext} and indeed showed that the cross-section ratio of $\sigma_{\Psi(2s)}/\sigma_{J/\Psi}$ exhibited weak dependence on dipole models, especially in the large $Q^2$ regime \cite{Chen:2016dlk}.  

\begin{widetext}

\begin{figure}
 \centering 
\includegraphics[width=.24\textwidth]{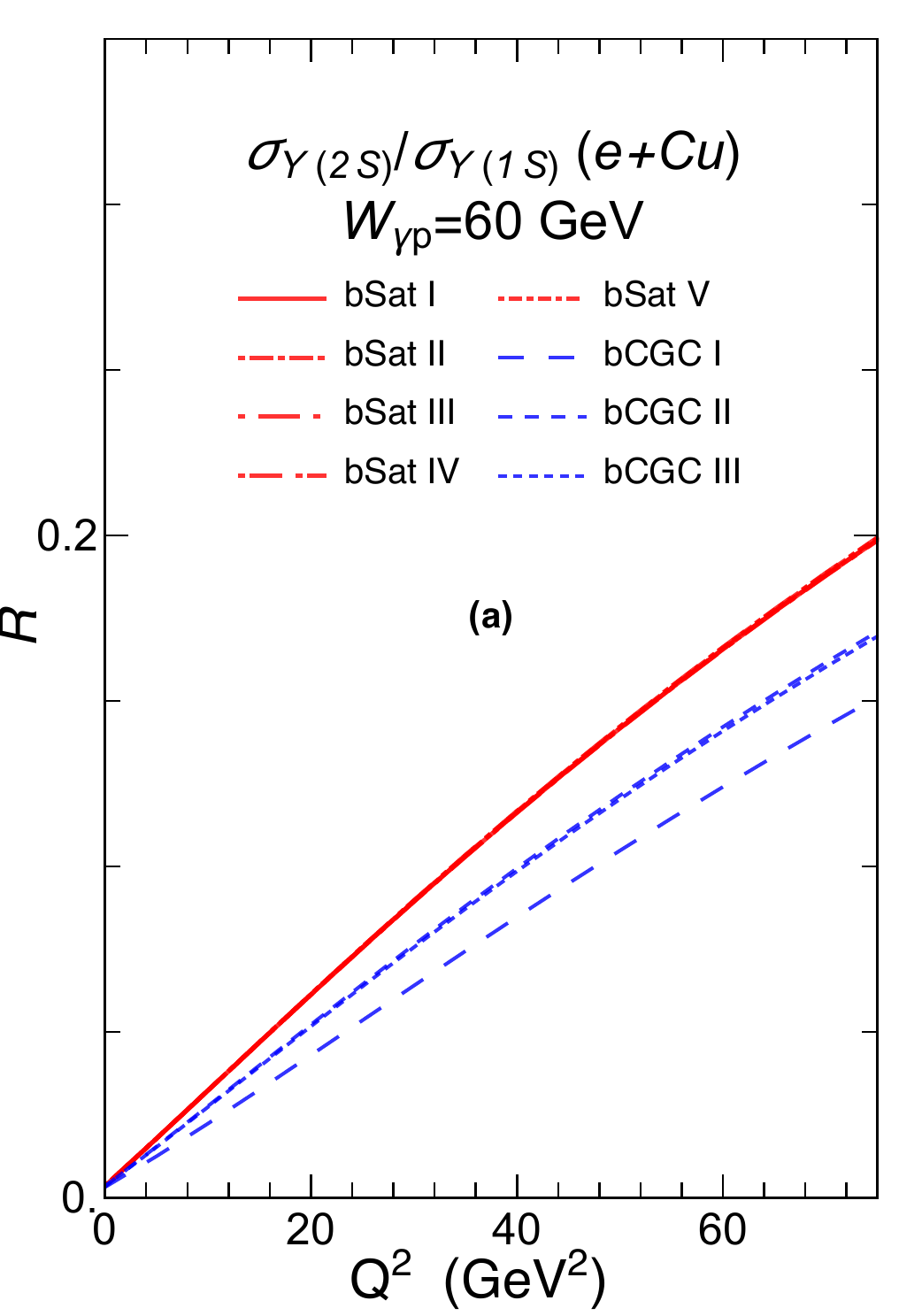}
\includegraphics[width=.24\textwidth]{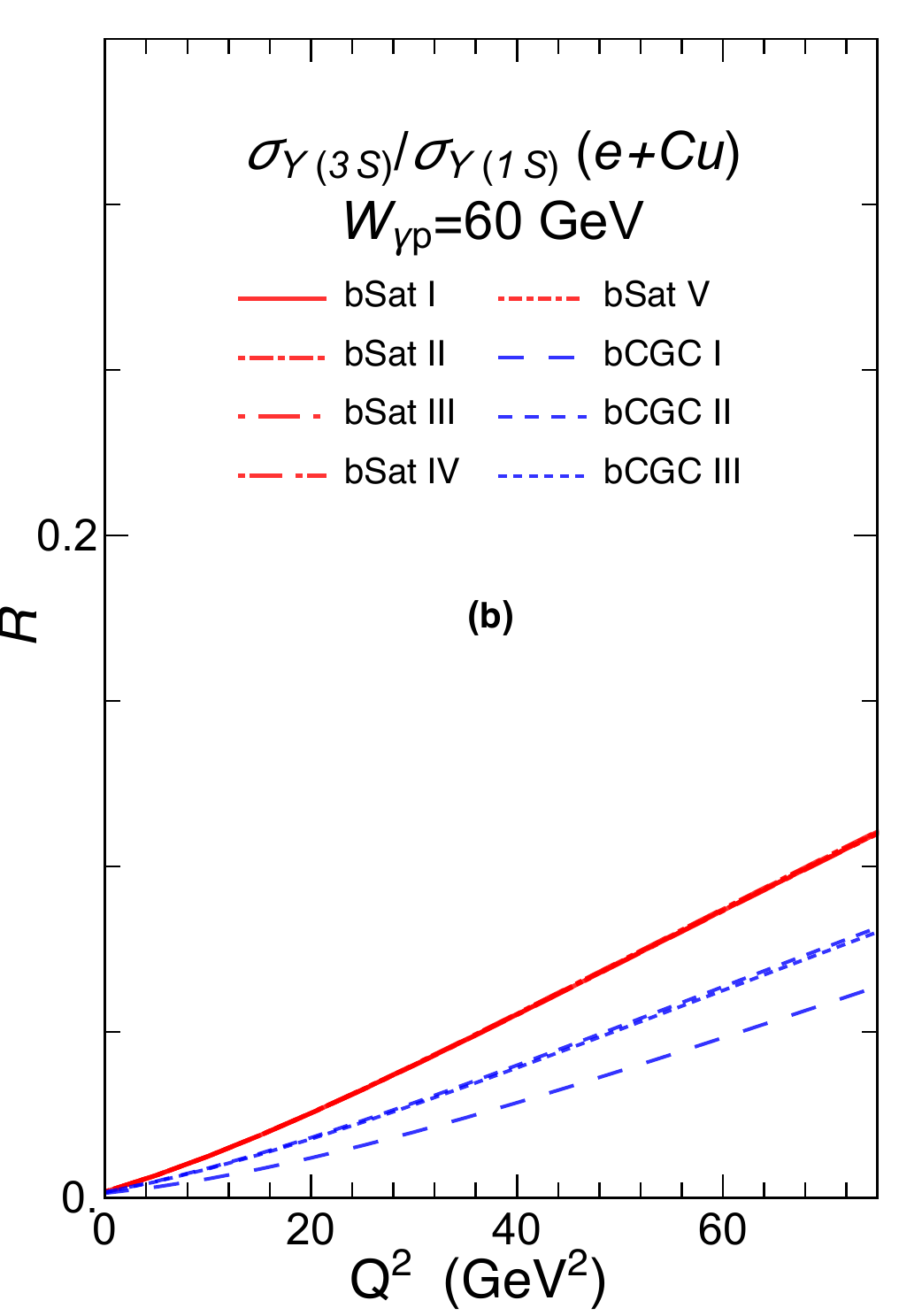}
\includegraphics[width=.24\textwidth]{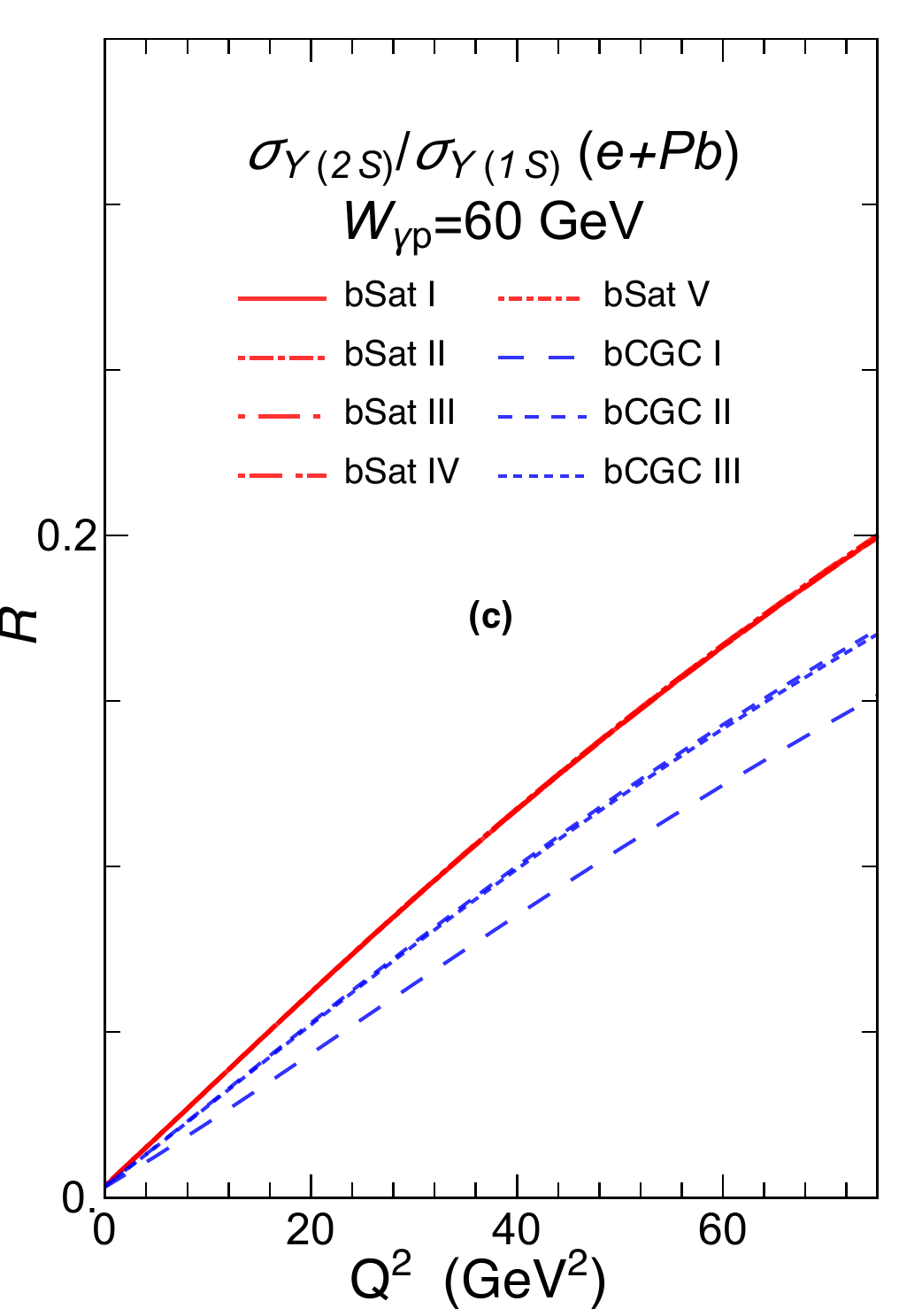}
\includegraphics[width=.24\textwidth]{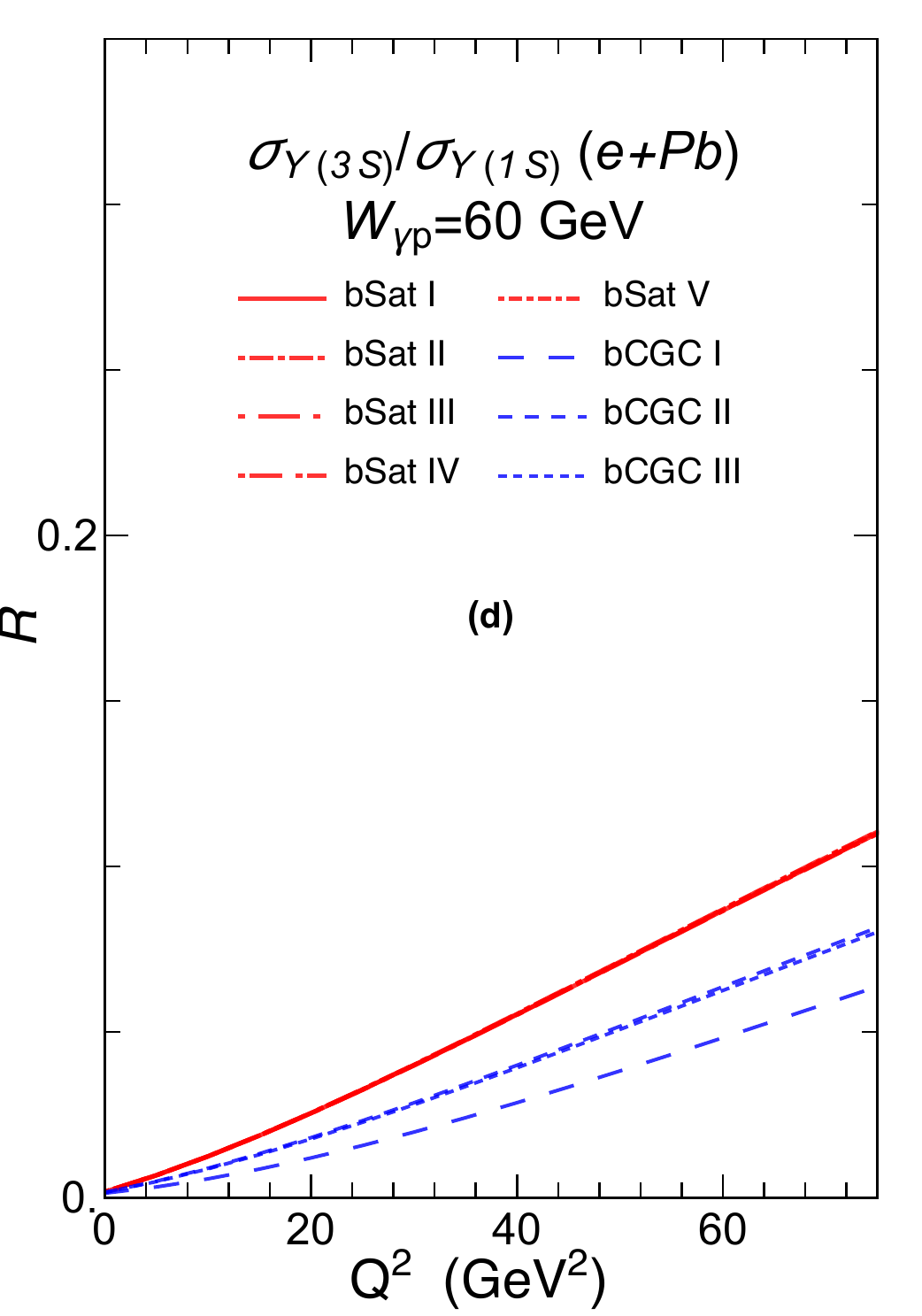}
\caption{(Colors online) The cross-section ratio $\sigma_{\Upsilon(2s)}/\sigma_{\Upsilon(1s)}$ and $\sigma_{\Upsilon(3s)}/\sigma_{\Upsilon(1s)}$ as a function of $Q^2$ predicted using the BLFQ LFWF using various dipole cross section parametrizations. The parameters of dipole models bSat I-V and bCGC I-III used in this calculation can be found in Table~1 and 2 in Ref.~\cite{Chen:2016dlk}. From left to right, we show the cross-section ratio $\sigma_{\Upsilon(2s)}/\sigma_{\Upsilon(1s)}$ (a) and $\sigma_{\Upsilon(3s)}/\sigma_{\Upsilon(1s)}$ (b) as a function of $Q^2$ in $e$Cu collision, and the cross-section ratio $\sigma_{\Upsilon(2s)}/\sigma_{\Upsilon(1s)}$ (c) and $\sigma_{\Upsilon(3s)}/\sigma_{\Upsilon(1s)}$ (d) as a function of $Q^2$ in $e$Pb collisions. Note that the predictions for several bSat parametrizations are almost indistinguishable from each other.}
\label{fig:model}
\end{figure}  

\end{widetext}
  
In this paper we calculate the cross-section ratios for Upsilon production in electron-ion collisions. In Fig.~\ref{fig:model}, we show the cross-section ratio $\sigma_{\Upsilon(2s)}/\sigma_{\Upsilon(1s)}$ and $\sigma_{\Upsilon(3s)}/\sigma_{\Upsilon(1s)}$ as a function of $Q^2$ predicted using the BLFQ LFWF with various dipole cross section parametrizations in $e$Cu collision and in $e$Pb collision. The parameters of dipole models bSat I-V and bCGC I-III used in this calculation can be found in Table~1 and 2 in Ref.~\cite{Chen:2016dlk}. We observe that the cross-section ratios for $\Upsilon$ states, exhibit weak dipole model dependence as well. Furthermore, the $\sigma_{\Upsilon(2s)}/\sigma_{\Upsilon(1s)}$ and $\sigma_{\Upsilon(3s)}/\sigma_{\Upsilon(1s)}$ cross-section ratios shows weak dependence on the colliding nucleus as well.  

A precise measurement of higher excited states of $\Upsilon$s is difficult at HERA or LHC but it would be achievable at a future Electron-Ion Collider \cite{Accardi:2012qut}. Our study in this paper and in Ref.~\cite{Chen:2016dlk} suggests that properties of the heavy quarkonium LFWFs could be investigated through measurements of cross-section ratios of higher excited states to the ground state in exclusive heavy quarkonium production, since the cross-section ratios are weakly dependent on the dipole model but show sensitivity to heavy quarkonium LFWFs. Furthermore, with well-constrained heavy quarkonium LFWFs, the three dimensional tomographic information of the proton can be extracted more efficiently from diffractive events.

\section{Conclusions and Outlook}
\label{sec:conclusion}

We study exclusive charmonium and bottomonium production in ultra-peripheral heavy-ion collisions and electron-ion collisions in the dipole picture. We employ heavy quarkonium light-front wavefunctions (LFWFs) obtained within the basis light-front quantization (BLFQ) approach. We have studied charmonium production using this BLFQ heavy quarkonium LFWFs and compared our predictions to selected experimental data at HERA, RHIC and LHC. Our results provide reasonably good descriptions of charmonium production for the experimental data we surveyed \cite{Chen:2016dlk}. In this investigation, we focus on comparing the theoretical prediction of the BLFQ LFWFs to experimental measurements of exclusive charmonium and bottomonium production in ultraperipheral $pp$, $p$Pb and PbPb collisions at LHC at various energies, including the data collected from run $2$ at LHC. For the new experimental data we surveyed, our theoretical predictions are in satisfactory agreement with experiment. Furthermore, we make predictions for the coherent production of $\Upsilon$s, including the excited states, at future electron-ion collision experiments. We find that the cross-section ratios of excited states to the ground state for bottomonium exhibit very little sensitivity to the dipole model parameters. Our study suggests that measuring such cross-section ratios at a future EIC could provide properties of the heavy quarkonium. A well-constrained heavy quarkonium wavefunction will be useful for extracting the $3$-dimensional tomographic information of the nucleon structure and will provide insights into gluon saturation physics.

The LFWFs we employed in this paper, which are obtained by diagonalizing an effective heavy quarkonium Hamiltonian in the BLFQ framework, have been found to provide reasonable descriptions of heavy quarkonia decay constants, radiative decay and form factors. We also show in this paper, consistent with a previous investigation, that the BLFQ LFWFs can also describe a wide range of experimental data for diffractive charmonium and bottomonium production. 

Our work indicates that the theoretical uncertainty is larger in the low $Q^2$ regime for the diffractive charmonium and bottomonium production. One possible improvement is to extend the BLFQ approach to higher Fock sectors, \eg, the quark-antiquark-gluon Fock sector, and use a dipole model parametrizations that incorporate the quark-antiquark-gluon sector as well. We believe such theoretical improvement would be a major advance for investigating diffractive heavy quarkonium production in future electron-ion collision experiments.

\acknowledgments

We thank P.~Maris and X.~Zhao for valuable disucssions. We also thank P.~Maris for his insightful comments and suggestions on the manuscript. This work was supported by the U.S. Department of Energy under Grant Nos. DE-FG02-87ER40371 and  DE-SC0018223 (SciDAC-4/NUCLEI). We acknowledge computational resources provided by the National Energy Research Scientific Computing Center (NERSC), which is supported by the Office of Science of the U.S. Department of Energy under Contract No. DE-AC02-05CH11231.



\begin{thebibliography}{99}
\bibitem{Gribov:1984tu} 
  L.~V.~Gribov, E.~M.~Levin and M.~G.~Ryskin,
  Phys.\ Rept.\  {\bf 100}, 1 (1983).
  
\bibitem{Derrick:1993xh} 
  M.~Derrick {\it et al.} [ZEUS Collaboration],
  Phys.\ Lett.\ B {\bf 315}, 481 (1993).
  
\bibitem{Ahmed:1994nw} 
  T.~Ahmed {\it et al.} [H1 Collaboration],
  Nucl.\ Phys.\ B {\bf 429}, 477 (1994).
  
\bibitem{Ivanov:2004ax} 
  I.~P.~Ivanov, N.~N.~Nikolaev and A.~A.~Savin,
  Phys.\ Part.\ Nucl.\  {\bf 37}, 1 (2006).
  
\bibitem{Wolf:2009jm} 
  G.~Wolf,
  Rept.\ Prog.\ Phys.\  {\bf 73}, 116202 (2010).
  
\bibitem{JalilianMarian:1996xn} 
  J.~Jalilian-Marian, A.~Kovner, L.~D.~McLerran and H.~Weigert,
  Phys.\ Rev.\ D {\bf 55}, 5414 (1997).
  
\bibitem{Golec-Biernat:1998js}
  K.~Golec-Biernat and M.~W\"usthoff,
  Phys.\ Rev.\ D {\bf 59}, 014017  (1999);
  K.~Golec-Biernat and M.~W\"usthoff,
  Phys.\ Rev.\ D {\bf 60}, 114023 (1999).
  
\bibitem{Levin:2000mv} 
  E.~Levin and K.~Tuchin,
  Nucl.\ Phys.\ A {\bf 691}, 779 (2001);
  E.~Levin and K.~Tuchin,
  Nucl.\ Phys.\ A {\bf 693}, 787 (2001).

\bibitem{Gotsman:2001ne} 
  E.~Gotsman, E.~Levin, U.~Maor and E.~Naftali,
  Phys.\ Lett.\ B {\bf 532}, 37 (2002).

\bibitem{Kowalski:2003hm} 
  H.~Kowalski and D.~Teaney,
  Phys.\ Rev.\ D {\bf 68}, 114005 (2003).


\bibitem{Iancu:2003ge}
  E.~Iancu, K.~Itakura and S.~Munier,
  Phys.\ Lett.\ B {\bf 590}, 199 (2004).
  
\bibitem{Gelis:2010nm} 
  F.~Gelis, E.~Iancu, J.~Jalilian-Marian and R.~Venugopalan,
  Ann.\ Rev.\ Nucl.\ Part.\ Sci.\  {\bf 60}, 463 (2010).
 
  
\bibitem{Mantysaari:2018nng} 
  H.~Mäntysaari and P.~Zurita,
  Phys.\ Rev.\ D {\bf 98}, 036002 (2018).
  
  
  
\bibitem{Abramowicz:2015mha} 
H.~Abramowicz {\it et al.} [H1 and ZEUS Collaborations],
  Eur.\ Phys.\ J.\ C {\bf 73}, no. 2, 2311 (2013);
H.~Abramowicz {\it et al.} [H1 and ZEUS Collaborations],
  Eur.\ Phys.\ J.\ C {\bf 75}, no. 12, 580 (2015).

\bibitem{AbelleiraFernandez:2012cc} 
  J.~L.~Abelleira Fernandez {\it et al.} [LHeC Study Group Collaboration],
  J.\ Phys.\ G {\bf 39}, 075001 (2012).

\bibitem{Accardi:2012qut} 
  A.~Accardi {\it et al.},
  Eur. Phys. J. A \textbf{52}, 268 (2016).

\bibitem{Mueller:1989st} 
  A.~H.~Mueller,
  Nucl.\ Phys.\ B {\bf 335}, 115 (1990).

\bibitem{Nikolaev:1990ja} 
  N.~N.~Nikolaev and B.~G.~Zakharov,
  Z.\ Phys.\ C {\bf 49}, 607 (1991).
    
\bibitem{Kopeliovich:1993pw} 
  B.~Z.~Kopeliovich, J.~Nemchick, N.~N.~Nikolaev and B.~G.~Zakharov,
  Phys.\ Lett.\ B {\bf 324}, 469 (1994).
  
\bibitem{Lappi:2016fmu} 
  T.~Lappi and H.~Mäntysaari,
  Phys.\ Rev.\ D {\bf 93}, no. 9, 094004 (2016).
  
\bibitem{Beuf:2016wdz} 
  G.~Beuf,
  Phys.\ Rev.\ D {\bf 94}, no. 5, 054016 (2016).
  
\bibitem{Nemchik:1994fp} 
  J.~Nemchik, N.~N.~Nikolaev and B.~G.~Zakharov,
  Phys.\ Lett.\ B {\bf 341}, 228 (1994).
  
\bibitem{Brodsky:1980vj}
  S.~J.~Brodsky, T.~Huang and G.~P.~Lepage,
  SLAC-PUB-2540.
  
  \bibitem{Nemchik:1996cw} 
  J.~Nemchik, N.~N.~Nikolaev, E.~Predazzi and B.~G.~Zakharov,
  Z.\ Phys.\ C {\bf 75}, 71 (1997).
  
\bibitem{Forshaw:2012im} 
  J.~R.~Forshaw and R.~Sandapen,
  Phys.\ Rev.\ Lett.\  {\bf 109}, 081601 (2012).
  
  
\bibitem{Cox:2009ag} 
  B.~E.~Cox, J.~R.~Forshaw and R.~Sandapen,
  JHEP {\bf 0906}, 034 (2009).

\bibitem{Vary:2009gt}
J.~P.~Vary, H.~Honkanen, J.~Li, P.~Maris, S.~J.~Brodsky, A.~Harindranath, G.~F.~de Teramond, P.~Sternberg, E.~G.~Ng and C.~Yang,
  Phys.\ Rev.\  C {\bf 81}, 035205 (2010). 

\bibitem{Honkanen:2010rc}
  H.~Honkanen, P.~Maris, J.~P.~Vary and S.~J.~Brodsky,
  Phys.\ Rev.\ Lett.\  {\bf 106}, 061603 (2011).


\bibitem{Zhao:2014xaa} 
  X.~Zhao, H.~Honkanen, P.~Maris, J.~P.~Vary and S.~J.~Brodsky,
  Phys.\ Lett.\ B {\bf 737}, 65 (2014).

\bibitem{Wiecki:2014ola} 
  P.~Wiecki, Y.~Li, X.~Zhao, P.~Maris and J.~P.~Vary,
  Phys.\ Rev.\ D {\bf 91}, 105009 (2015) .
  
\bibitem{Vary:2016emi} 
  J.~P.~Vary, L.~Adhikari, G.~Chen, Y.~Li, P.~Maris and X.~Zhao,
  Few Body Syst.\  {\bf 57}, no. 8, 695 (2016).

%
 
\bibitem{deTeramond:2005su} 
  G.~F.~de Teramond and S.~J.~Brodsky,
  Phys.\ Rev.\ Lett.\  {\bf 102}, 081601 (2009).
  
\bibitem{Brodsky:2014yha} 
  S.~J.~Brodsky, G.~F.~de Teramond, H.~G.~Dosch and J.~Erlich,
  Phys.\ Rept.\  {\bf 584}, 1 (2015).
  
\bibitem{Li:2015zda} 
  Y.~Li, P.~Maris, X.~Zhao and J.~P.~Vary,
  Phys.\ Lett.\ B {\bf 758}, 118 (2016). Tables of the wavefunctions from this paper are available at http://dx.doi.org/10.17632/5bgp37xwz4.1.

\bibitem{Li:2017mlw} 
  Y.~Li, P.~Maris and J.~P.~Vary,
  Phys.\ Rev.\ D {\bf 96}, no. 1, 016022 (2017).

\bibitem{Li:2018uif} 
  M.~Li, Y.~Li, P.~Maris and J.~P.~Vary,
  Phys.\ Rev.\ D {\bf 98}, no. 3, 034024 (2018).
  
\bibitem{Chen:2016dlk} 
  G.~Chen, Y.~Li, P.~Maris, K.~Tuchin and J.~P.~Vary,
  Phys.\ Lett.\ B {\bf 769}, 477 (2017).


  
\bibitem{Rezaeian:2013tka} 
  A.~H.~Rezaeian and I.~Schmidt,
  Phys.\ Rev.\ D {\bf 88}, 074016 (2013).

\bibitem{Kowalski:2006hc} 
  H.~Kowalski, L.~Motyka and G.~Watt,
  Phys.\ Rev.\ D {\bf 74}, 074016 (2006).
  
\bibitem{Lepage:1980fj} 
  G.~P.~Lepage and S.~J.~Brodsky,
  Phys.\ Rev.\ D {\bf 22}, 2157 (1980).

\bibitem{Forshaw:2003ki}
  J.~R.~Forshaw, R.~Sandapen and G.~Shaw,
  Phys.\ Rev.\ D {\bf 69}, 094013 (2004).
  
\bibitem{Nikolaev:1994rd} 
  N.~N.~Nikolaev, B.~G.~Zakharov and V.~R.~Zoller,
  J.\ Exp.\ Theor.\ Phys.\  {\bf 78}, 806 (1994).

\bibitem{Mueller:1994jq} 
  A.~H.~Mueller and B.~Patel,
  Nucl.\ Phys.\ B {\bf 425}, 471 (1994).

\bibitem{BFKL}   
L.N. Lipatov, Sov.\ J.\ Nucl.\ Phys.\  {\bf 23}, 338 (1976); E.A. Kuraev,  L.N. Lipatov and V.S. Fadin, Sov.\ Phys.\ JETP {\bf 44},
443 (1976);
E.A. Kuraev, L.N. Lipatov and V.S. Fadin,Sov.\ Phys.\ JETP {\bf 28}, 822 (1978).

\bibitem{BK}     
I. Balitsky, 
Nucl.\ Phys.\ B {\bf 463}, 99 (1996);
Y. V. Kovchegov, 
Phys.\ Rev.\ D {\bf 60}, 034008 (1999).

\bibitem{Levin:1999mw} 
  E.~Levin and K.~Tuchin,
  Nucl.\ Phys.\ B {\bf 573}, 833 (2000).
  
\bibitem{Watt:2007nr} 
  G.~Watt and H.~Kowalski,
  Phys.\ Rev.\ D {\bf 78}, 014016 (2008).
  
\bibitem{Bartels:2002cj} 
  J.~Bartels, K.~J.~Golec-Biernat and H.~Kowalski,
  Phys.\ Rev.\ D {\bf 66}, 014001 (2002).
  
\bibitem{Rezaeian:2012ji} 
  A.~H.~Rezaeian, M.~Siddikov, M.~Van de Klundert and R.~Venugopalan,
  Phys.\ Rev.\ D {\bf 87}, 034002 (2013).
  
  
    
   
  
%
%
  
\bibitem{Ryskin:1995hz} 
  M.~G.~Ryskin, R.~G.~Roberts, A.~D.~Martin and E.~M.~Levin,
  Z.\ Phys.\ C {\bf 76}, 231 (1997).
  
%
%


\bibitem{Brodsky:2008pg} 
  S.~J.~Brodsky and G.~F.~de Teramond,
  Subnucl.\ Ser.\  {\bf 45}, 139 (2009).

\bibitem{Zhao:2013cma} 
  X.~Zhao, A.~Ilderton, P.~Maris and J.~P.~Vary,
  Phys.\ Rev.\ D {\bf 88}, 065014 (2013).
  
\bibitem{Chen:2017uuq} 
  G.~Chen, X.~Zhao, Y.~Li, K.~Tuchin and J.~P.~Vary,
  Phys.\ Rev.\ D {\bf 95}, no. 9, 096012 (2017).
  
  
\bibitem{Baltz:2007kq} 
  A.~J.~Baltz {\it et al.},
  Phys.\ Rept.\  {\bf 458}, 1 (2008).
  
  
\bibitem{Bertulani:2005ru} 
  C.~A.~Bertulani, S.~R.~Klein and J.~Nystrand,
  Ann.\ Rev.\ Nucl.\ Part.\ Sci.\  {\bf 55}, 271 (2005).
  
\bibitem{Kopeliovich:2001ee} 
  B.~Kopeliovich, A.~Tarasov and J.~Hufner,
  Nucl.\ Phys.\ A {\bf 696}, 669 (2001).

\bibitem{Goncalves:2014wna} 
  V.~P.~Goncalves, B.~D.~Moreira and F.~S.~Navarra,
  Phys.\ Rev.\ C {\bf 90}, 015203 (2014).
  
\bibitem{Xie:2016ino} 
  Y.~P.~Xie and X.~Chen,
  Eur.\ Phys.\ J.\ C {\bf 76}, no. 6, 316 (2016).
  
\bibitem{Goncalves:2017wgg} 
  V.~P.~Gonçalves, M.~V.~T.~Machado, B.~D.~Moreira, F.~S.~Navarra and G.~S.~dos Santos,
  Phys.\ Rev.\ D {\bf 96}, no. 9, 094027 (2017).
  
\bibitem{Carvalho:2017vtw} 
  F.~Carvalho, V.~P.~Goncalves, F.~S.~Navarra and D.~Spiering,
  Phys.\ Rev.\ D {\bf 97}, no. 7, 074002 (2018).
  
\bibitem{Mantysaari:2017dwh} 
  H.~Mäntysaari and B.~Schenke,
  Phys.\ Lett.\ B {\bf 772}, 832 (2017).
  
\bibitem{Dutta:2017kju} 
  D.~Dutta and R.~Chudasama,
  arXiv:1711.05999 [hep-ph].
  
  
  
  
  
  
  
\bibitem{Aaij:2014iea} 
  R.~Aaij {\it et al.} [LHCb Collaboration],
  J.\ Phys.\ G {\bf 41}, 055002 (2014).

\bibitem{Aaij:2018arx} 
  R.~Aaij {\it et al.} [LHCb Collaboration],
  arXiv:1806.04079 [hep-ex].


\bibitem{Aaij:2015kea} 
  R.~Aaij {\it et al.} [LHCb Collaboration],
  JHEP {\bf 1509}, 084 (2015).
  
\bibitem{TheALICE:2014dwa} 
  B.~B.~Abelev {\it et al.} [ALICE Collaboration],
  Phys.\ Rev.\ Lett.\  {\bf 113}, no. 23, 232504 (2014).
  
\bibitem{Kryshen:2017jfz} 
  E.~L.~Kryshen [ALICE Collaboration],
  Nucl.\ Phys.\ A {\bf 967}, 273 (2017).
  
\bibitem{Abbas:2013oua} 
  E.~Abbas {\it et al.} [ALICE Collaboration],
  Eur.\ Phys.\ J.\ C {\bf 73}, 2617 (2013).

\bibitem{Khachatryan:2016qhq} 
  V.~Khachatryan {\it et al.} [CMS Collaboration],
  Phys.\ Lett.\ B {\bf 772}, 489 (2017).

\bibitem{LHCb:2018ofh} 
  [LHCb Collaboration],
  LHCb-CONF-2018-003, CERN-LHCb-CONF-2018-003.
  
\bibitem{Abramowicz:2016ext} 
  H.~Abramowicz {\it et al.} [ZEUS Collaboration],
Nucl.\ Phys.\ B {\bf 909}, 934 (2016).






%




%



  
  
    
\end{thebibliography}
\end{document}